\documentclass[11pt,a4paper]{article}
\pdfoutput=1
\usepackage{graphicx}
\usepackage{jheppub}
\usepackage{cancel}

  \usepackage{amsmath}
  \usepackage{amsfonts}
  \usepackage{amssymb}

  \usepackage{graphicx}
  \usepackage{float}
\newcommand{\ba}{\begin{eqnarray}}
\newcommand{\ea}{\end{eqnarray}}

\title{Simulations of Cold Electroweak Baryogenesis: Hypercharge U(1) and the creation of helical magnetic fields}

	\author[b]{Zong-Gang Mou,}
	\author[a]{Paul M. Saffin,}
	\author[b]{Anders Tranberg}

	\affiliation[a]{School of Physics and Astronomy, University Park, University of Nottingham,\\ Nottingham NG7 2RD, United Kingdom}
	\affiliation[b]{Faculty of Science and Technology, University of Stavanger, 4036 Stavanger, Norway}

	\emailAdd{zonggang.mou@uis.no}
	\emailAdd{paul.saffin@nottingham.ac.uk}
	\emailAdd{anders.tranberg@uis.no}
	

\abstract{We perform numerical simulations of Cold Electroweak Baryogenesis, including for the first time in the Bosonic sector the full electroweak gauge group SU(2)$\times$U(1) and CP-violation. We find that the maximum generated baryon asymmetry is reduced by a factor of three relative to the SU(2)-only model of \cite{CBquench}, but that the quench time dependence is very similar. In addition, we compute the magnitude of the helical magnetic fields, and find that it is proportional to the strength of CP-violation and dependent on quench time, but is not proportional to the magnitude of the baryon asymmetry as proposed in \cite{tanmay1,tanmay2}. Astrophysical signatures of primordial magnetic helicity can therefore not in general be used as evidence that electroweak baryogenesis has taken place.}

\begin{document}

\maketitle

\section{Introduction}
\label{sec:Intro}

In Cold Electroweak Baryogenesis, the out-of-equilibrium conditions responsible for creating the baryon asymmetry of the Universe originate from a fast quench of the Higgs potential at zero temperature. Rather than the first order thermal phase transition and bubble nucleation of Hot Electroweak Baryogenesis (realised in some extensions of the Standard Model
), the spinodal instability ensures exponential growth of IR modes of the Higgs field, which in turn drive energy into the gauge fields. In the presence of CP-violation, this straightforwardly produces a baryon asymmetry \cite{Krauss,GB,Copeland,Jan1}. This asymmetry depends linearly on the magnitude of CP-violation and, it turns out, has a maximum for a finite quench time \cite{Jan_quench,CBquench}.

A strong appeal of this scenario is that it is amenable to essentially first-principles numerical field theory simulation and computation of the baryon asymmetry. But these simulations are numerically intensive, as the equations to be solved are often implicit, the physical volumes to be simulated must be large to include the IR dynamics and it is necessary to average over an ensemble of hundreds of classical random realisations.

Previous simulations of Cold Electroweak Baryognesis with CP-violation have included only the SU(2) gauge group of the Standard Model (in some cases with a second Higgs field \cite{wu1,wu2}  or with fermions \cite{fermions}). Several simulations of the CP-even dynamics have also considered hypercharge and the generation of magnetic fields \cite{juan1,juan2}, or have included an additional (inflaton) field \cite{GarciaBellido:2003wd}, presumed to be the trigger of the symmetry breaking transition.

We will here revisit this scenario, but including both hypercharge U(1) as well as CP-violation. This allows us to compute the expectation value of both CP-even observables (Higgs field, magnetic field, electric field, energy components, variance of CP-odd observables) \cite{juan1,juan2}, but also CP-odd ones (helical magnetic field, Chern-Simons numbers, Higgs winding number). From the CP-even observables, we will be able to track the process of tachyonic preheating, and how the energy is transferred to the degrees of freedom present, on the short and long term. 
Of particular interest to us is the effect of including hypercharge on the final and maximal baryon asymmetry created, and discovering how the asymmetry depends on the quench time. Also, we will investigate whether ``secondary" CP-odd observables such as helicity of the magnetic field also acquire a non-zero expectation value, and find to what extent it reflects the CP-violation strength and quench time dependence of the baryon asymmetry itself. 

It has been conjectured, that in electroweak baryogenesis, the amplitude of the net helicity should be proportional to the net baryon asymmetry of the Universe \cite{tanmay1,tanmay2,juan2} (see also \cite{Grasso:2000wj}). Helicity is expected to be conserved in time as a result of the very large conductivity of the ambient plasma in the Universe. This implies that observation of such helicity would be an indication of electroweak dynamics being responsible for baryogenesis. Conversely, the observed galactic and intergalactic magnetic fields could be explained by the electroweak phase transition. We will argue that this conjecture is based on features specific to hot electroweak baryogenesis and the decay of sphaleron configurations, and does not apply to Cold Electroweak Baryogenesis. We will demonstrate this numerically. 

The paper is structured as follows: In section \ref{sec:model} we will present our SU(2)$\times$U(1)-Higgs model with CP-violation, and present the set of observables we will compute.  In section 
\ref{sec:CPeven} we will consider CP-even observables and their short and long-time behaviour. In section \ref{sec:CPodd} we compute the CP-odd observables and study their dependence on the strength of CP-violation and quench time, as well as the relationship between helical magnetic field and Chern-Simons number, and hence baryogenesis. We conclude in section \ref{sec:conc}.

\section{The SU(2)$\times$U(1)-Higgs model with CP-violation}
\label{sec:model}

We model the electroweak-sector of the Standard Model by a Higgs doublet coupled to SU(2) and U(1) gauge fields, with the classical action
\begin{eqnarray}\label{eq:action}
S&=&- \int dt\, d^3x\Bigg[ \frac{1}{2}\textrm{Tr}\,W^{\mu\nu}W_{\mu\nu} +\frac{1}{4} B^{\mu\nu}B_{\mu\nu}+ (D^\mu\phi)^\dagger D_\mu\phi \nonumber\\
	&~&\qquad\qquad\qquad+\mu^2_{\rm eff}(t)\phi^\dagger\phi + \lambda(\phi^\dagger\phi)^2 +\frac{3\delta_{\rm cp}g^2}{16\pi^2 m_W^2}\phi^\dagger\phi \textrm{Tr}\,W^{\mu\nu}\tilde{W}_{\mu\nu}\Bigg].\nonumber\\
\end{eqnarray}
The field strength tensors are $W_{\mu\nu}$ for SU(2) and $B_{\mu\nu}$ for U(1), with the dual defined by \mbox{$\tilde{W}_{\mu\nu}=\frac{1}{2}\epsilon_{\mu\nu\rho\sigma}W^{\rho\sigma}$}.  The gauge couplings are $g$ and $g'$, respectively, and we have a Higgs self-interaction $\lambda$, in addition to a time-dependent mass-coefficient $\mu_{\rm eff}(t)$ defined by
\ba
\mu_{\rm eff}^2(t)&=&\left\{
\begin{array}{cc}
\mu^2(1-2t/\tau_q) &\qquad\qquad t<\tau_q,\\
-\mu^2                   & \qquad\qquad t\geq\tau_q.
\end{array}
\right.
\ea
It is initially positive, ``restoring symmetry", and once the quench is completed, \mbox{$\mu_{\rm eff}^2(t)\rightarrow -\mu^2$}, so that the Standard Model vacuum expectation value satisfies $\lambda v^2=\mu^2$. 

The covariant derivative $D_\mu$ is given by
\begin{eqnarray}
D_\mu\phi = \left(\partial_\mu +i \frac{1}{2}g'B_\mu-i gW_\mu^a\frac{\sigma_a}{2}\right)\phi,
\end{eqnarray}
with the U(1) gauge field $B_\mu$ and the SU(2) gauge field denoted by $W_\mu$. We have used that the Higgs field hypercharge is $Y=-1/2$.

The discrete CP-symmetry is broken by the last term of (\ref{eq:action}) because $W^{\mu\nu}\tilde{W}_{\mu\nu}$ breaks P, preserving C. For successful electroweak baryogenesis, it is necessary to break P, C and CP separately, so that baryon number (C-odd) and Chern-Simons number (P-odd) can become non-zero. For the full Standard Model-like theory with fermions, C and P are broken by the gauge-fermion coupling, while the present term breaks P. 
$\delta_{\rm cp}$ is a dimensionless measure of the strength of CP-violation. In the simulations it will be taken in the range 0-7, with the physical baryon asymmetry corresponding to values around $\delta_{\rm cp}\simeq 10^{-5}$ \cite{Jan_kappa,CBquench}.  

The origin of the CP-violating term is assumed to be interactions at a higher energy scale, possibly involving heavy fermions. We take it as a generic higher order operator with the required symmetry (breaking) properties. Integrating out the fermions of the Standard Model itself produces effective terms similar to this, although not exactly the same \cite{CPSM}. The equivalent of $\delta_{\rm cp}$ for the Standard Model is a complicated expression in the parameters of the CKM mass matrix and the temperature, and turns out to be orders of magnitude too small to account for the baryon asymmetry \cite{CPSM}, unless the effective temperature is around \mbox{$T=1$ GeV}. 

Our choice of P-breaking term biases the time derivative of the SU(2) Chern-Simons number, defined as an integral over time
\begin{eqnarray}
\label{eq:Ncstime}
N_{\rm cs,2}(t)-N_{\rm cs,2}(0) = \frac{g^2}{16\pi^2}\int_0^t dt\, d^3x\, \textrm{Tr}\, W^{\mu\nu}\tilde{W}_{\mu\nu},
\end{eqnarray}
or equivalently in a spatial representation
\begin{eqnarray}
\label{eq:Ncsspace}
N_{\rm cs,2}(t) = -\frac{g^2}{32\pi^2}\int d^3 x\, \epsilon^{ijk}\left(W_i^a W_{jk}^a-\frac{g}{3}\epsilon_{abc}W_i^aW_j^bW_k^c\right).
\end{eqnarray}
When this bias manifests itself as a non-zero expectation value of the Chern-Simons number, the baryon asymmetry follows from the anomaly equation
\begin{eqnarray}
\label{eq:anomaly}
\langle N_B(t)-N_B(0)\rangle = 3\langle\left[N_{\rm cs,2}(t)-N_{\rm cs,2}(0)\right]\rangle.
\end{eqnarray}
We will ultimately infer the baryon asymmetry from the final value of the Higgs winding number
\begin{eqnarray}
N_{\rm w}= \frac{1}{24\pi^2}\int dx^3 \epsilon_{ijk}\textrm{Tr}[U^\dagger\partial_i UU^\dagger\partial_j UU^\dagger\partial_k U], 
\end{eqnarray}
with 
\begin{eqnarray}
U(x)=\frac{1}{\phi^\dagger\phi} (i\tau_2\phi^*,\phi).
\end{eqnarray}
At late times, $N_{\rm w}\simeq N_{\rm cs}$ is enforced dynamically, but since winding number settles early in the simulation, it is more convenient to use from a computer-time perspective. 
The hypercharge field also provides a CP-odd observable, that we will refer to as the U(1) Chern-Simons number
\begin{eqnarray}\label{eq:u1ChernSimons}
N_{\rm cs,1}(t)= -\frac{(g')^2}{32\pi^2}\int d^3 x\, \epsilon^{ijk}B_i B_{jk},
\end{eqnarray}
in terms of the spatial components of the gauge field and its field strength. This is not a Chern-Simons number in the sense of a winding number, but it does enter in the anomaly equation for fermion number. Strictly speaking, eq. (\ref{eq:anomaly}) should be \begin{eqnarray}
\langle N_B(t)-N_B(0)\rangle = 3\langle\left[(N_{\rm cs,2}-N_{\rm cs,1})(t)-(N_{\rm cs,2}-N_{\rm cs,1})(0)\right]\rangle,
\end{eqnarray}
but since only SU(2) has a non-trivial set of equivalent vacua, a vacuum-to-vacuum transition involving true fermion production always has $\Delta N_{\rm cs,1}=0$.

Additional observables that will be of interest include the average Higgs field 
\begin{eqnarray}
\bar{\phi}^2=\frac{1}{V}\int d^3x\,\phi^\dagger\phi,
\end{eqnarray}
and the various energy components in SU(2), U(1) and Higgs fields, respectively,  $E_W$, $E_B$, $E_\phi$. 

As symmetry breaking proceeds, the Higgs mechanism promotes different linear combinations of field degrees of freedom to be mass eigenstates. The $Z$ boson and the photon field $A$ are defined in the unitary gauge in the broken phase as
\begin{eqnarray}
Z_\mu = W^3_\mu \cos\theta + B_\mu \sin\theta,\qquad A_\mu = W^3_\mu \sin\theta - B_\mu \cos\theta.
\end{eqnarray}
These are naturally generalised to
\begin{eqnarray}
Z_\mu = n^aW^a_\mu \cos\theta + B_\mu \sin\theta,\qquad A_\mu = n^aW^a_\mu \sin\theta - B_\mu \cos\theta,
\end{eqnarray}
outside of the unitary gauge, where we have defined
\begin{eqnarray}
n^a=-\varphi^\dagger \sigma^a\varphi,\qquad\varphi= \frac{\phi}{|\phi|}.
\end{eqnarray}
However, due to the ambiguity in the definition of the photon \cite{tHooft:1974kcl}, which is only really defined in the symmetry-broken phase, we find it convenient to add a term proportional to 
\mbox{$\varphi^\dagger\partial_\mu\varphi-\partial_\mu\varphi^\dagger\varphi$} to our definition of the electromagnetic field, allowing us to write it in a manifestly gauge covariant way as
\begin{eqnarray}
A_\mu= \frac{i}{g}\left[(D_\mu\varphi)^\dagger\varphi-\varphi^\dagger(D_\mu\varphi)
\right]\sin\theta- \frac{B_\mu}{ \cos\theta},
\end{eqnarray}
with the field strength defined as the curl in the usual way, leading to
\begin{eqnarray}\label{eq:emFieldStrength}
F_{\mu\nu}=\left[\partial_\mu(n^aW^a_\nu)-\partial_\nu(n^aW^a_\mu)-\frac{1}{g}\epsilon_{abc}n^a\partial_\mu n^b \partial_\nu n^c
\right]\sin\theta-B_{\mu\nu}\cos\theta.
\end{eqnarray}
By construction, in the unitary gauge in the broken phase, $n^a\rightarrow (0,0,1)$ everywhere, and the expression simplifies to the standard expressions. However, we will perform our simulation in temporal gauge $A_0=0$, and therefore we need a gauge covariant expression that is simple to use on a lattice. We note that since we will be using the zero temperature value for the mixing angle $\theta$, the field $A_\mu$ is the photon field throughout, but only at the end of the transition is it massless \cite{sphaleron}. We will include in our list of observables the magnetic energy component associated with the photon field, $E_{mag}$. This is not a distinct component, being composed of part of $E_B$ and part of $E_W$.

Once the symmetry breaking transition is complete, we have for the masses of the system
\begin{eqnarray}
m_H &=& 2\mu^2 = \sqrt{2\lambda} v=125 \textrm{ GeV},\\
m_W&=&\frac{1}{2}g v=77.5 \textrm{ GeV},\\
m_Z &=&\frac{m_W}{\cos\theta}= 88.4 \textrm{ GeV},\\
m_\gamma &=&0,
\end{eqnarray}
with our choices of parameters\footnote{We use primarily low energy-scale values for the couplings, since the transition starts out at zero temperature. Using only electroweak-scale parameters amounts to a correction of a few percent and has no impact on the final results.}
\begin{eqnarray}
v=246 \textrm{ GeV},\qquad
\sin^2\theta = 0.231,\qquad
e=\sqrt{\frac{4\pi}{137}},\qquad
g= \frac{e}{\sin \theta}=0.63,\qquad
g'= \frac{e}{\cos\theta}=0.35.\nonumber\\
\end{eqnarray}

In the $A-Z$ field basis, we define the helical magnetic field of the photon field to be
\begin{eqnarray}
N_h =  -\frac{1}{2}\int d^3 x \epsilon^{ijk}A_i F_{jk}.
\end{eqnarray}
Note that this has the same structure as the Chern-Simons number of the U(1) hypercharge field that was defined previously (\ref{eq:u1ChernSimons}).
Since the photon field has contributions from both SU(2) and U(1) gauge fields, we may expect some correlation with $N_{\rm cs, 2}$ and/or $N_{\rm cs, 1}$ \cite{tanmay1,tanmay2}. 

To summarize, our CP-odd (P-odd, C-even) observables are $N_{\rm w}$, $N_{\rm cs, 1}$, $N_{\rm cs, 2}$ and $N_h$, while our CP-even (P-even, C-even) observables are $\phi^2$, all the energy components $E_B$, $E_W$, $E_\phi$, $E_{mag}$ as well as the squares of the CP-odd observables. 

We discretize the system on a spatial lattice, and solve the classical equations of motion for an explicitly CP-symmetric ensemble of random initial conditions. The initial conditions for the Higgs field are chosen to mimic a quantum vacuum prior to the mass quench \cite{half_Paul,juan3,Jan_q}. The gauge fields are initially zero, with the gauge momenta found from Gauss's Law in the background of the initial Higgs field. Statistical errors are computed based on the differences between CP-conjugate initial condition pairs. We average over ensembles of 100-300 such pairs. The lattice size is $Vm_H^3=24^3$, with $64^3$ lattice sites and $am_H=0.375$. This is large enough to include the IR physics, while still keeping lattice artefacts under control. The total numerical effort for the data presented here is in the order of 100.000 cpu hours. Details on the numerical implementation can be found elsewhere \cite{Jan1}, while explicit expressions for the lattice equations of motion and observables may be found in Appendix \ref{sec:lattice}. 

\section{CP-even averages: The preheating process and real-time rates}
\label{sec:CPeven}

\subsection{Flips}
\label{sec:flips}

\begin{figure}
\begin{tabular}{lr}
\hspace{-1cm}
\includegraphics[width = 0.55\textwidth]{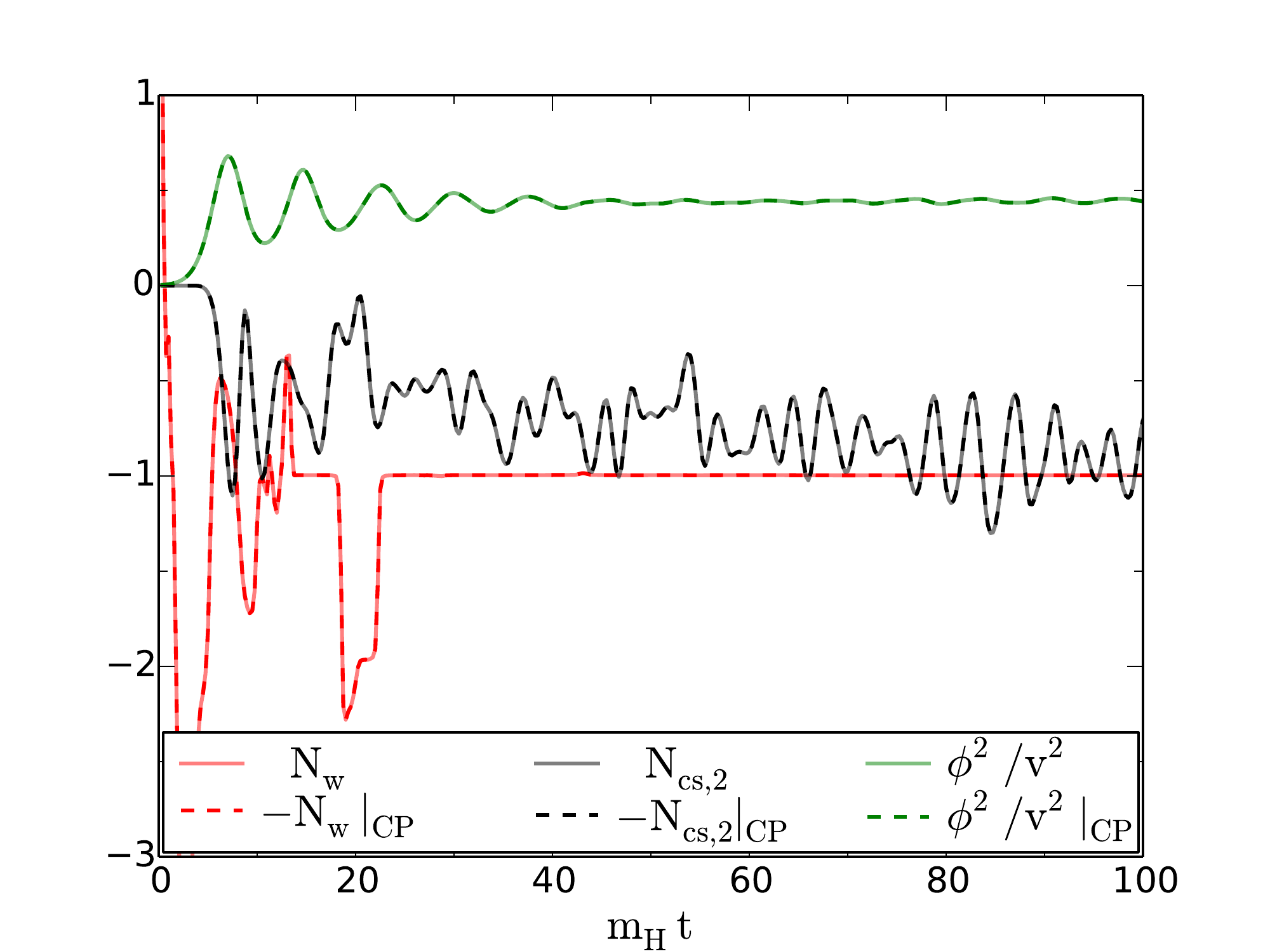} & \hspace{-1.2cm} 
\includegraphics[width = 0.55\textwidth]{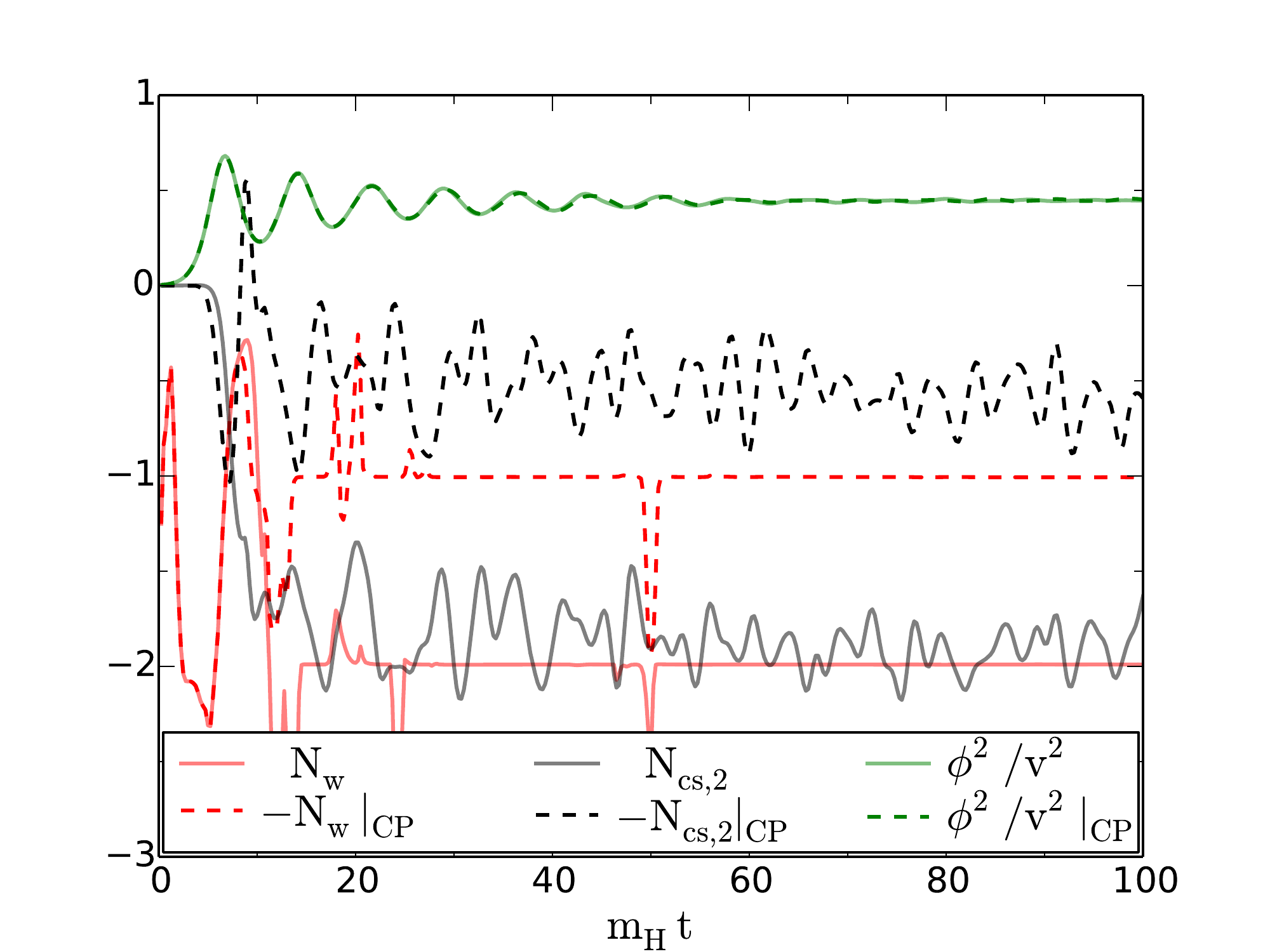}
\end{tabular}
\caption{Some observables for single random realisations of the initial conditions. Left: Without CP-violation, no flip. Right: With CP-violation, showing an example where a flip occurs. The notation $N_{\rm w}$ and $\left.N_{\rm w}\right|_{\rm cp}$, for example, refers to the evolutions from a random initial condition and the evolution from the CP-conjugated initial data.}
 \label{fig:Even_single}
\end{figure}

As symmetry breaking is triggered by $\mu_{\rm eff}^2(t)$ becoming negative the Higgs field IR modes become unstable. The Higgs field effectively ``rolls down" from its potential maximum, and energy is transferred to the gauge fields as particles are created. At first, only the modes $k \leq |\mu_{\rm eff}(t)|$ grow exponentially, but as non-linear interactions kick in, the energy is redistributed into the UV modes as well \cite{Skullerud:2003ki}. As kinetic equilibration completes over a timescale of a few hundred in mass units, the spectrum acquires an approximate exponential form, similar to a Bose-Einstein distribution with an effective chemical potential. Chemical equilibrium then shifts the distribution over a timescale an order of magnitude longer. The redistribution stage has features similar to turbulence or a cascade \cite{juan1,juan2, Skullerud:2003ki}.

In Fig. \ref{fig:Even_single} (left) we show, for a single random configuration pair \footnote{ A configuration pair is the two simulations whose initial data is related by a CP transformation.}, the time history of $\phi^2$, $N_{\rm cs,2}$ and $N_{\rm w}$. For the CP-odd observables, we have flipped the sign of one of the two configurations, so that they appear on top of each other, when they are precisely of opposite sign. We see the spinodal roll-off of the Higgs field turn into damped oscillations, while the CP-odd observables evolve into non-zero values. For these simulations, the CP-violation parameter $\delta_{\rm cp}$ was set to zero, but for individual configurations CP-odd observables can still be non-zero. In fact, the figure shows the trajectories for a CP-conjugate pair, and we see that the two have exactly opposite values of the CP-odd observables ($N_{\rm cs,1}$ and $N_h$ are not shown but behave similarly). They average to zero identically in our CP-symmetric ensemble composed of such pairs.

In Fig. \ref{fig:Even_single} (right) we show another pair of configurations, with CP-violation turned on. We see that the CP-odd observables are opposite until time $10\,m_H^{-1}$, where they diverge, giving a non-zero integer average over the ensemble. We call such an event a ``flip" (in this case a ``-1 flip"). Most configurations do not flip, even with  CP-violation present, but a certain fraction flip with $\pm 1$ or even in rare cases $\pm 2$. Averaging over these instances of $0,\pm 1,\pm 2$ gives our ensemble averages. We find, that when CP-violation is not present, the vanishing of the CP-odd observables is not a result of the cancelling of positive and negative flips. No flips occur at all. 

\subsection{Energy}
\label{sec:energy}

\begin{figure}
\centering
    \includegraphics[width = 0.65\textwidth]{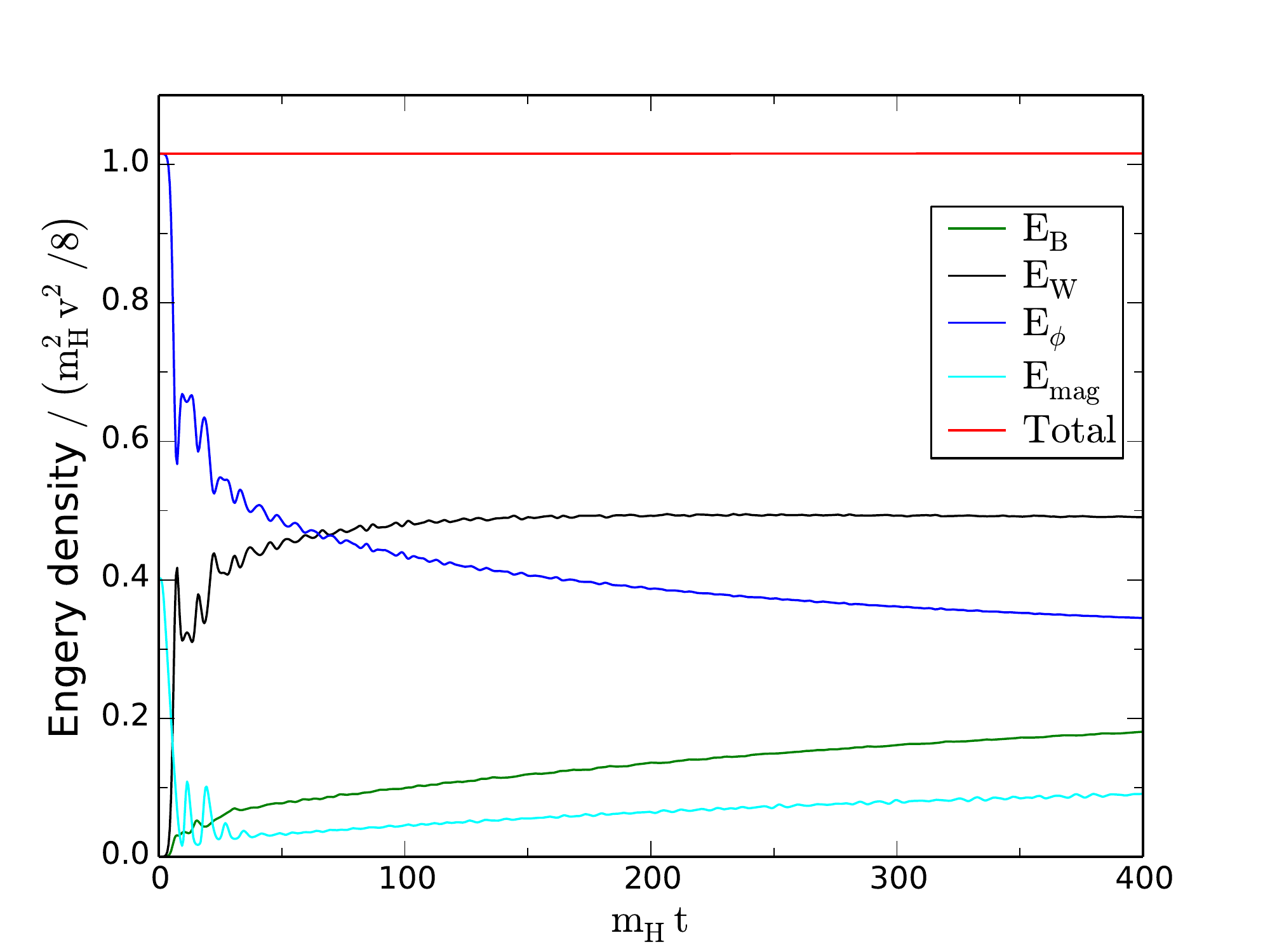}
\caption{The different energy components, averaged over an explicitly CP-symmetric ensemble, for an instantaneous quench $\tau_q=0$.}
 \label{fig:E_average}
\end{figure}

 In Fig. \ref{fig:E_average}, we show the different energy components $E_\phi$, $E_{B}$ and $E_{W}$, corresponding to the three fields in the problem. In addition, we show the magnetic energy density $E_{mag}$, defined in the usual way, taking (\ref{eq:emFieldStrength}) as the field strength. We see that as the spinodal transition happens, energy pours into these dynamical degrees of freedom. The SU(2) energy and also U(1) energy grow very rapidly initially, during the first few Higgs oscillations.  This, in effect, is the process of post-inflationary preheating for this scenario. Afterwards the energy starts equipartitioning between the various degrees of freedom. Being not in unitary gauge, the precise counting of degrees of freedom between field variables is non-trivial, but at the latest time shown here, the energy fraction is roughly $E_W:E_\phi:E_B=6:4:2$, suggestive of effectively massless gauge bosons coupled to four Higgs degrees of freedom.

\begin{figure}
\centering
    \includegraphics[width = 0.65\textwidth]{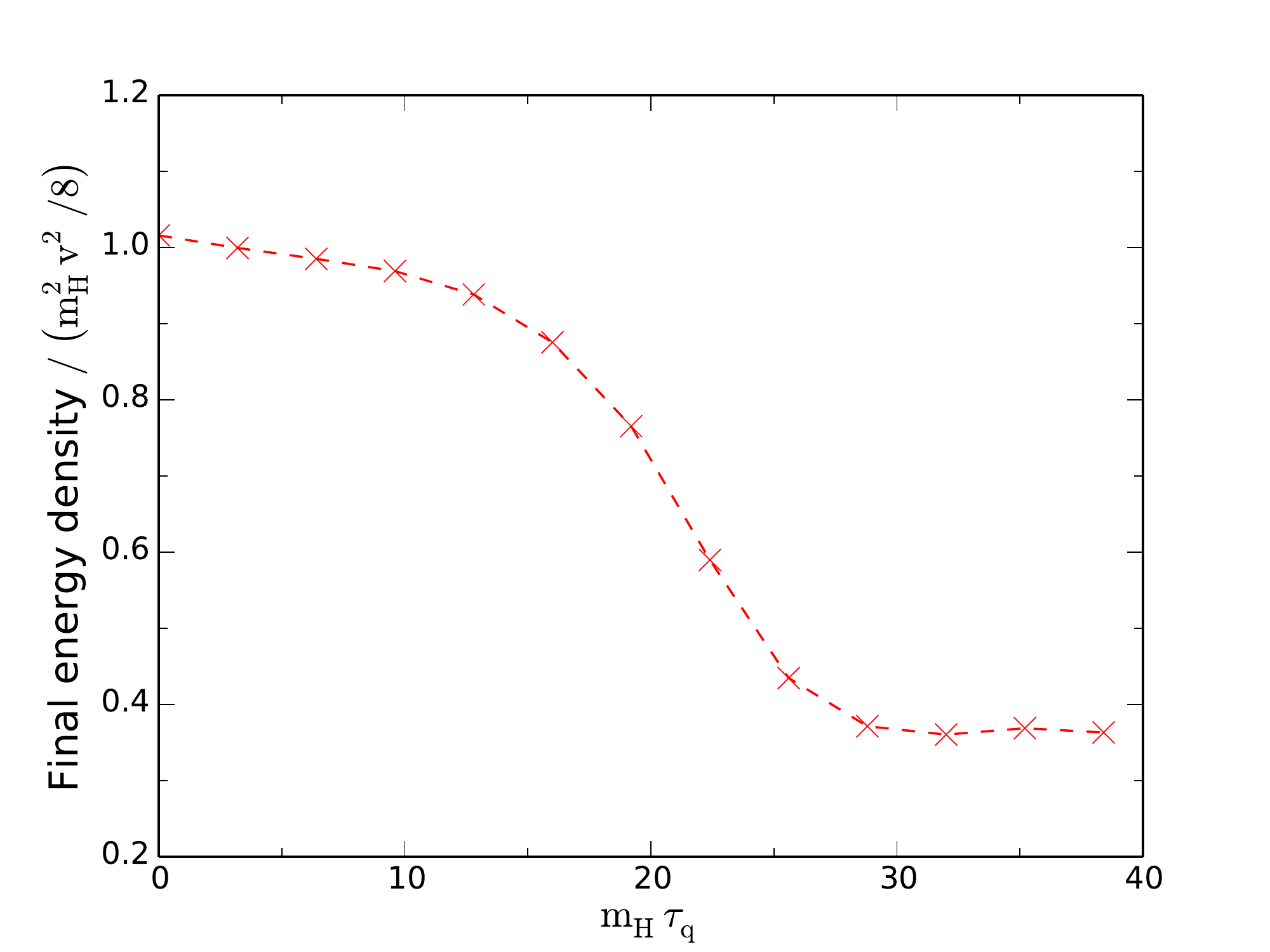}
\caption{The final energy density as a function of quench time.}
 \label{fig:deltaE}
\end{figure}

An issue with our implementation, is that because $\mu_{\rm eff}(t)$ is not a dynamical variable, but an external ``source", energy is not conserved in the system. The energy loss can be written as
\begin{eqnarray}
\Delta E = -\frac{2\mu^2}{\tau_q}\int_0^{\tau_q} dt\, d^3x\,\phi^\dagger\phi(x,t),
\end{eqnarray}
and computed numerically, as shown in Fig.~\ref{fig:deltaE}, as a function of quench time. We see that for our quench times, as much as $\frac{2}{3}$ of the energy is extracted in this way. This means that the energy available to preheat the fields is reduced for slow quenches. The final temperature is also smaller by a factor $T_{\rm slow quench}\propto (E_{\rm initial}-\Delta E)^{1/4}\simeq 0.7 \,T_{\rm fast quench}$. By redistribution of the initial potential energy to a thermal state at late times, this corresponds to temperatures of about $70-50$ GeV, counting all the degrees of freedom in the model as massless. Either way, this is deep in the ``broken symmetry" phase of the Standard Model. We point out that the energy extraction enters through the Higgs equation of motion only, and not the gauge fields themselves. 

For a complete model, where the quench is triggered by a dynamical field (an inflaton or curvaton, or some other scalar \cite{Copeland,mod1,mod2,mod3}), no energy is lost, although part of it will end up in this additional degree of freedom \cite{juan1,juan2}. The field dynamics however also becomes more complicated. We will address this in a future publication \cite{inflaton_us}, and proceed here with a hand-made quench, keeping in mind the energy loss for slow quenches. 

\subsection{Variance of observables}
\label{sec:variance}

\begin{figure}
\begin{tabular}{lr}
\hspace{-1cm}
\includegraphics[width = 0.55\textwidth]{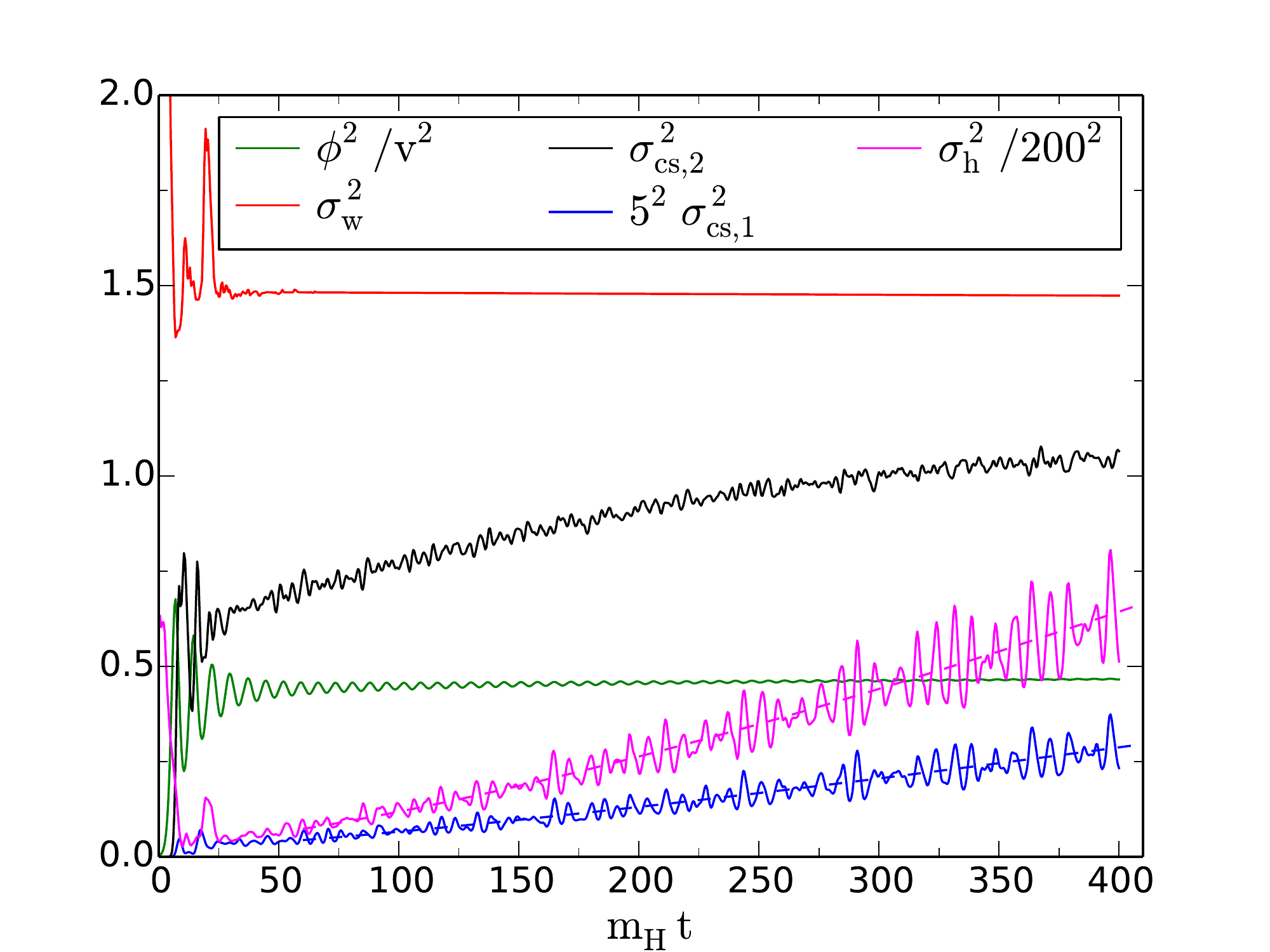} & \hspace{-1.2cm} 
\includegraphics[width = 0.55\textwidth]{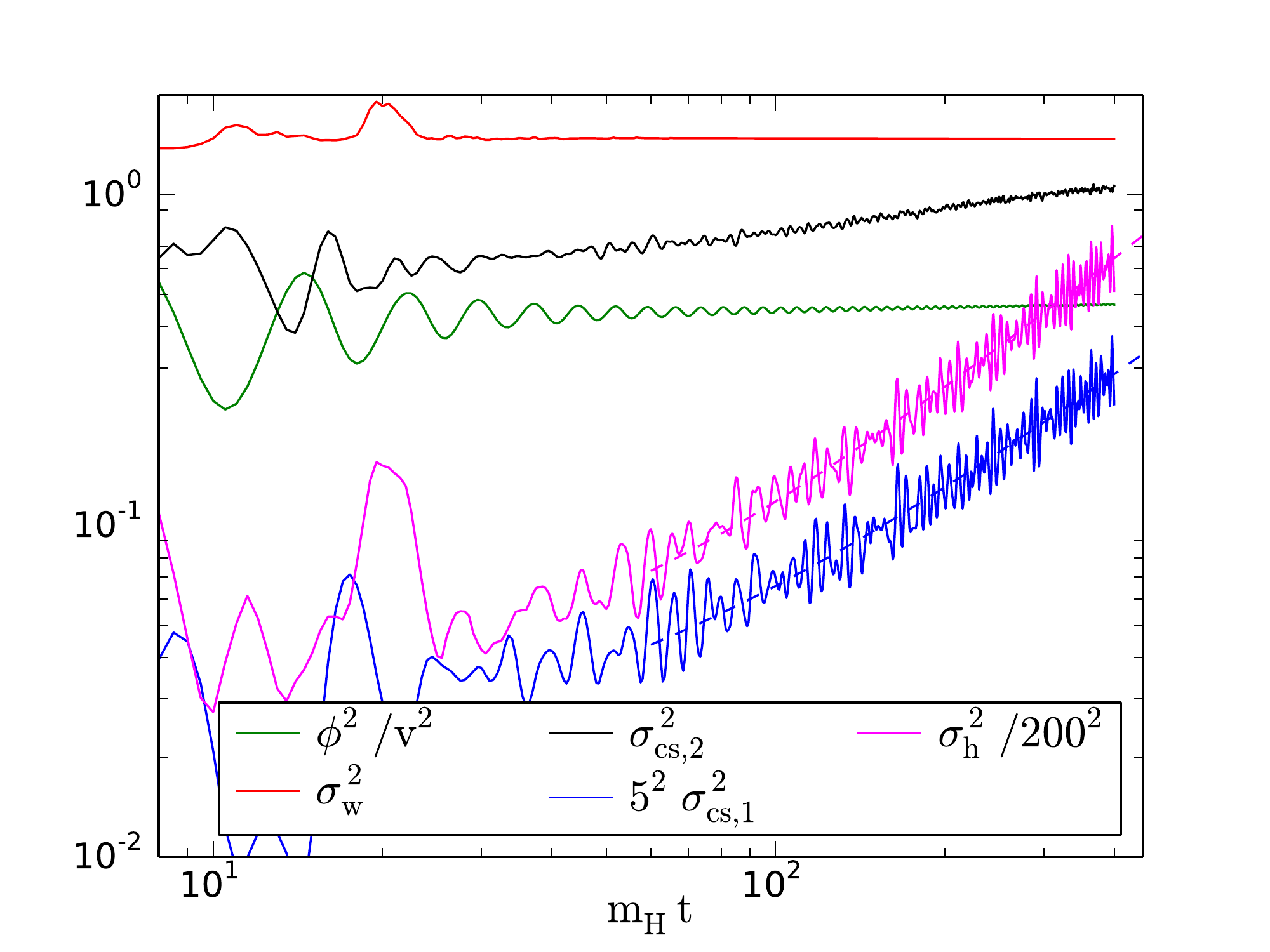}
\end{tabular}
\caption{The variance of the CP-odd observables, averaged over an explicitly CP-symmetric ensemble. We use $\delta_{\rm cp}=6.83$, and the quench is instantaneous $\tau_q=0$. The helicity and U(1) Chern-Simons number have been rescaled to fit inside the plot.}
\label{fig:Even_average}
\end{figure}
One effect of preheating is that as the fields get populated and temperature increases, the variance of various observables also increases. We are interested in the square of the CP-odd quantities
\begin{eqnarray}
\label{eq:square_obs}
\sigma_{\rm cs,2}^2&=&\langle N_{\rm cs,2}^2\rangle-\langle N_{\rm cs,2}\rangle^2,\quad \sigma_{\rm cs,1}^2=\langle N_{\rm cs,1}^2\rangle-\langle N_{\rm cs,1}\rangle^2,\\\nonumber
\sigma_{\rm w}^2&=&\langle N_{\rm w}^2\rangle-\langle N_{\rm w}\rangle^2,\quad \sigma_h^2=\langle N_h^2\rangle-\langle N_h\rangle^2,
\end{eqnarray}
from which one may compute the time derivatives, the ``diffusion rates"
\begin{eqnarray}
\Gamma_{\rm cs,2}=\frac{d\sigma_{\rm cs,2}^2}{dt},\quad
\Gamma_{\rm cs,1}=\frac{d\sigma_{\rm cs,1}^2}{dt},\quad
\Gamma_{\rm w}=\frac{d\sigma_{\rm w}^2}{dt},\quad
\Gamma_{\rm h}=\frac{d\sigma_h^2}{dt}.
\end{eqnarray}
In equilibrium, the first of these is the familiar Sphaleron rate \cite{Khlebnikov:1988sr,Burnier:2005hp} (up to a factor of the volume), describing the finite temperature diffusion of Chern-Simons number. In our case, we are not in equilibrium, and the observable is a priori just a time-dependent rate of change of Chern-Simons number squared. Similarly for the other observables. 

In Fig. \ref{fig:Even_average} (left, linear and right, log scale), we show the observables (\ref{eq:square_obs}), averaged over an  ensemble, for various quench times with $\delta_{\rm cp}=6.83$. We observe that the winding number settles completely around time $m_Ht\simeq 50$, as it becomes fixed in each configuration, fixing the average and the variance. The other observables continue to increase also at later times. At asymptotically late times, the rate is given by the equilibrium diffusion rate at the final temperature. Since this temperature is in the range 50-75 GeV, at least for Chern-Simons number the Sphaleron rate is exponentially suppressed. We see this as an approach to a constant value of the variance. 
In contrast, we see 
that the variance of $N_{\rm cs,1}$ and $N_h$ at late times grow as a power law, 
with late time dependence found to be $\sim(m_Ht)^{1.4}$ and $\sim(m_Ht)^{1.3}$ for $\sigma^2_h$ and $\sigma^2_{\rm cs,1}$ respectively, at the physical Higgs mass $m_H=1.6 \,m_{\rm w}$. 

One may have hoped to be able to estimate the asymmetry in each of the CP-odd observables, created by the introduction of a CP-breaking bias, without explicitly doing the simulations. A ``linear response"-like treatment was proposed in \cite{Jan_quench} for the Chern-Simon number as
\begin{eqnarray}
\label{eq:boltz}
\langle N_{\rm cs, 2}\rangle(t) = \int \frac{\mu_{\rm cs,2}(t)}{T_{\rm eff}(t)}\Gamma_{\rm cs, 2}(t) dt,
\end{eqnarray}
where we have generalized the expression by allowing a time-dependent chemical potential $\mu_{\rm cs,2}$ to provide a CP-bias, and allow for a time-dependent effective temperature $T_{\rm eff}$. Such an expression may of course be written down for any CP-odd observable, in terms of each their diffusion rate, and bias coefficient. 

For instance in the present case of using (\ref{eq:action}) and (\ref{eq:Ncstime}), we may readily find that the P violating term in the action is
\begin{eqnarray}
S_{\cancel{\rm p}}&\sim&\int dt\;\mu_{\rm cs,2}(t) N_{\rm cs},\qquad\mu_{\rm cs,2}(t)=\frac{3\delta_{\rm cp}}{m_{\rm w}^2}\frac{d}{{d}t}\langle\phi^\dagger\phi\rangle,
\end{eqnarray}
where we have approximated the quadratic scalar piece by its spatial average. This shows that the time dependence of the Higgs vev gives a chemical potential to the Chern-Simons number, and so we may write a Boltzmann equation of the form (\ref{eq:boltz}).

\begin{figure}
\begin{tabular}{lr}
\hspace{-1cm}
\includegraphics[width = 0.55\textwidth]{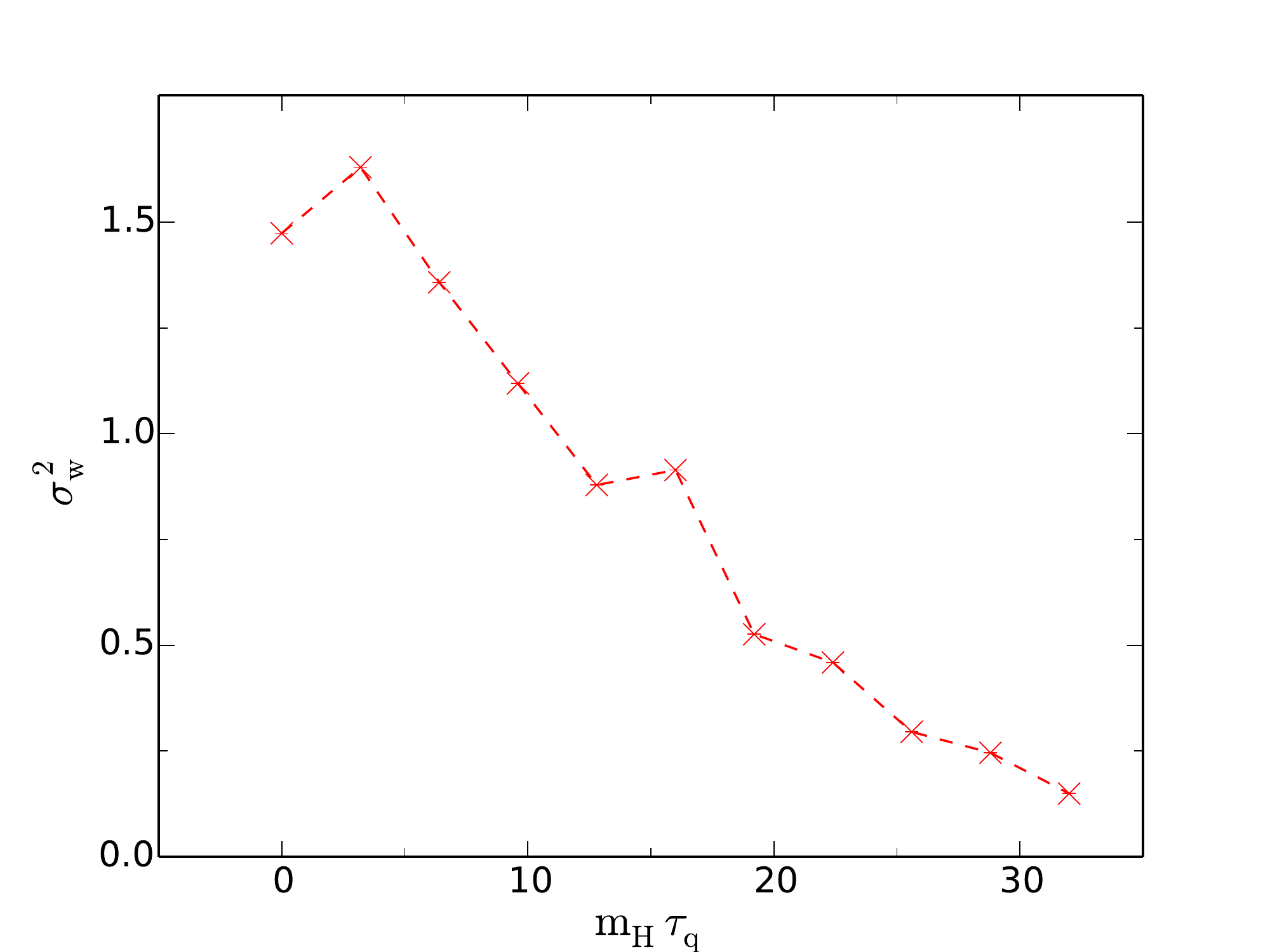} & \hspace{-1.2cm} 
\includegraphics[width = 0.55\textwidth]{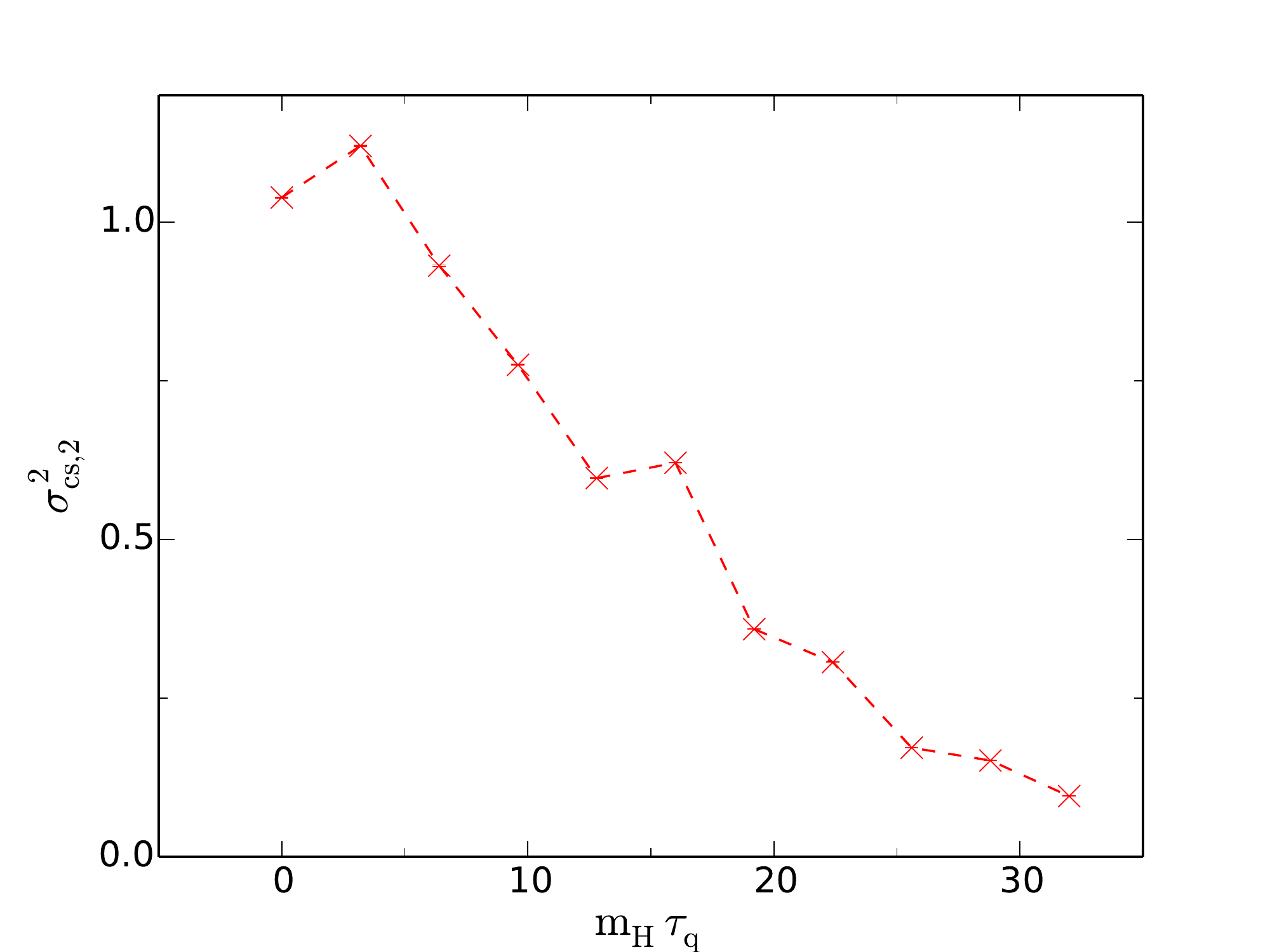} \\
\hspace{-1cm}
\includegraphics[width = 0.55\textwidth]{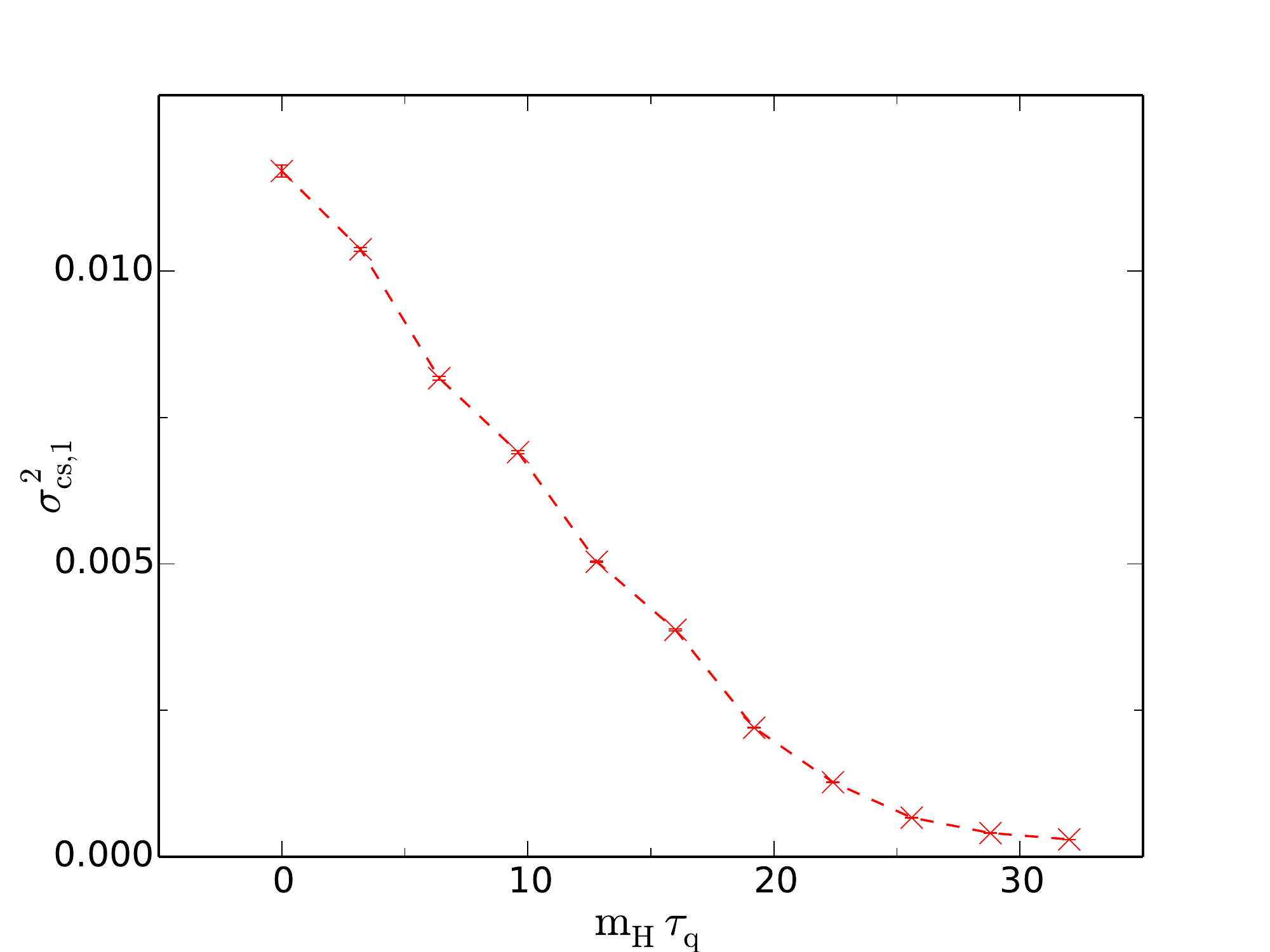} & \hspace{-1.2cm} 
\includegraphics[width = 0.55\textwidth]{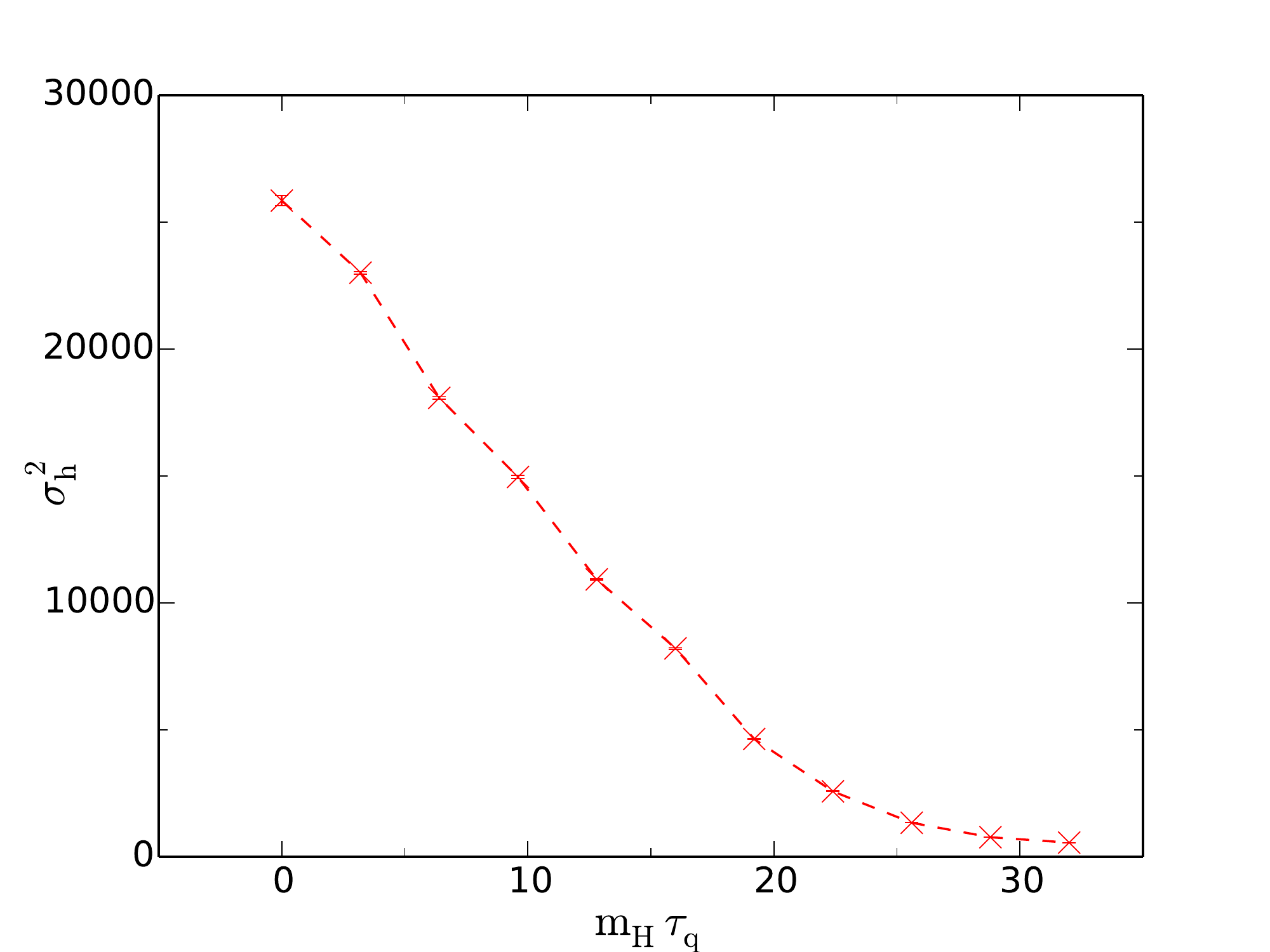} 
\end{tabular}
\caption{The final value of the variances as a function of quench time, for $\delta_{\rm cp}=6.83$.}
 \label{fig:Even_quench}
\end{figure}

It was found in \cite{Jan_quench}, that only the very initial stages of the baryogenesis process could be reproduced with such a treatment, and then only by tuning the value of the effective temperature $T_{\rm eff}$. This shows that the asymmetry is not driven by diffusion. For instance, we note that if  $\mu_{\rm cs,2}/T_{\rm eff}$ is taken to be constant, the final asymmetry follows from integrating up the diffusion rate, and simply find
\begin{eqnarray}
\langle N_{\rm cs,2}\rangle(t) =\frac{\mu_{\rm cs,2}}{T_{\rm eff}}\langle N_{\rm cs,2}^2\rangle(t).
\end{eqnarray}
In particular, since the variances are always positive, the overall sign of any asymmetry would be given by the sign of $\delta_{\rm cp}$. In Fig.~\ref{fig:Even_quench} we show the values of the observables squared at late times\footnote{We evaluate the late-time variances at $390\,m_H^{-1}$ after the first minimum of $\bar\phi^2$, and we find that the first minimum occurs at $0.71m_H\tau_q+10$.} for different quench times. We see that they are all monotonically decreasing.
We found that even allowing for quite unrealistic time-dependence of $\mu_{\rm cs,2/cs,1/h}/T_{\rm eff}$ we were unable to reproduce the behaviour of any of the CP-odd observables.

\section{CP-odd: The baryon asymmetry and helicity}
\label{sec:CPodd}

\subsection{Short and long time evolution}
\label{sec:shortlong}

\begin{figure}
\centering
    \includegraphics[width = 0.65\textwidth]{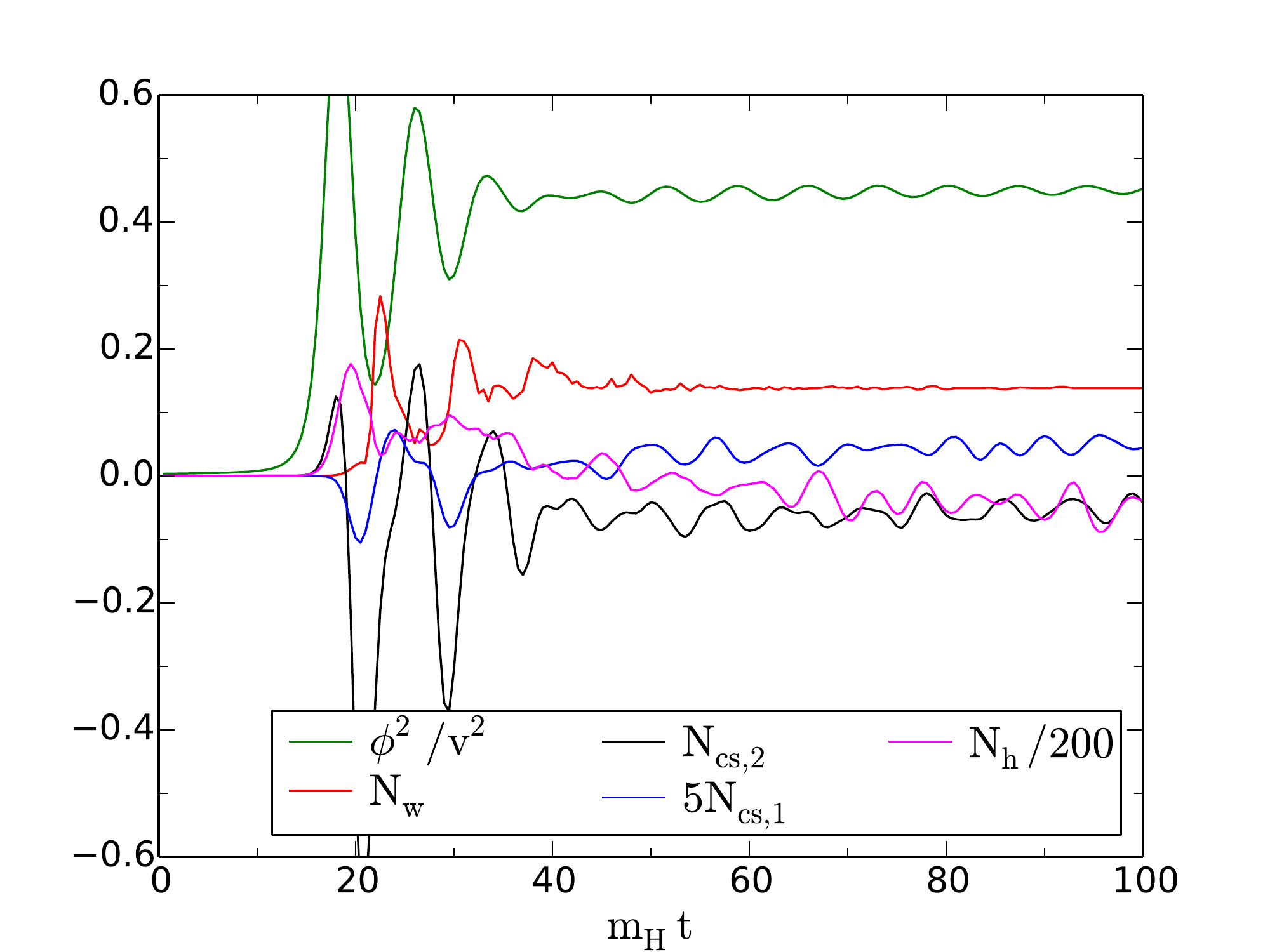}
\caption{The ensemble averaged observables for $m_H\tau_q=16$ and $\delta_{\rm cp}=6.83$. }
 \label{fig:Ex_averages}
\end{figure}

\begin{figure}
\begin{tabular}{lr}
\hspace{-1cm}
\includegraphics[width = 0.55\textwidth]{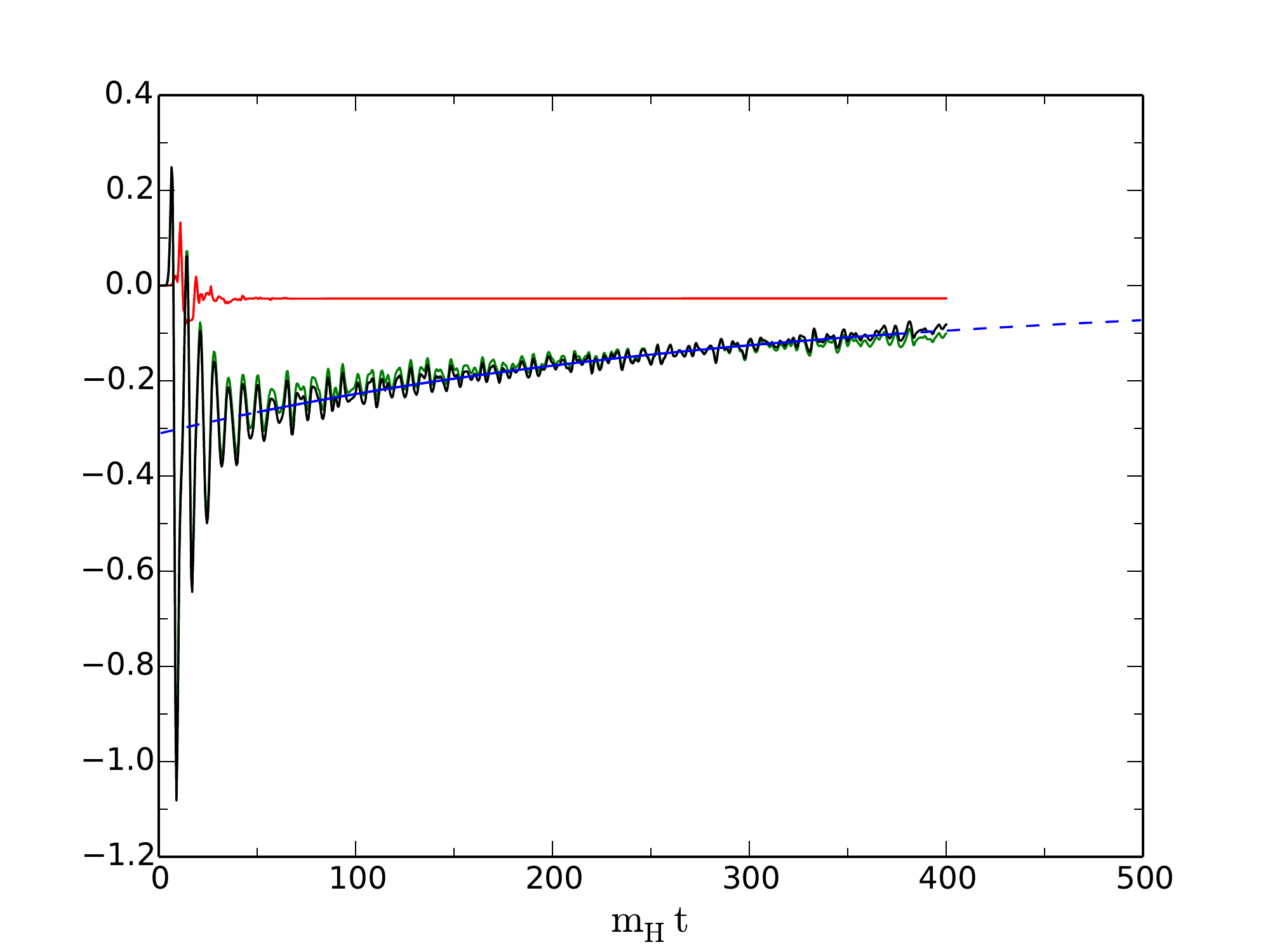} & \hspace{-1.2cm} 
\includegraphics[width = 0.55\textwidth]{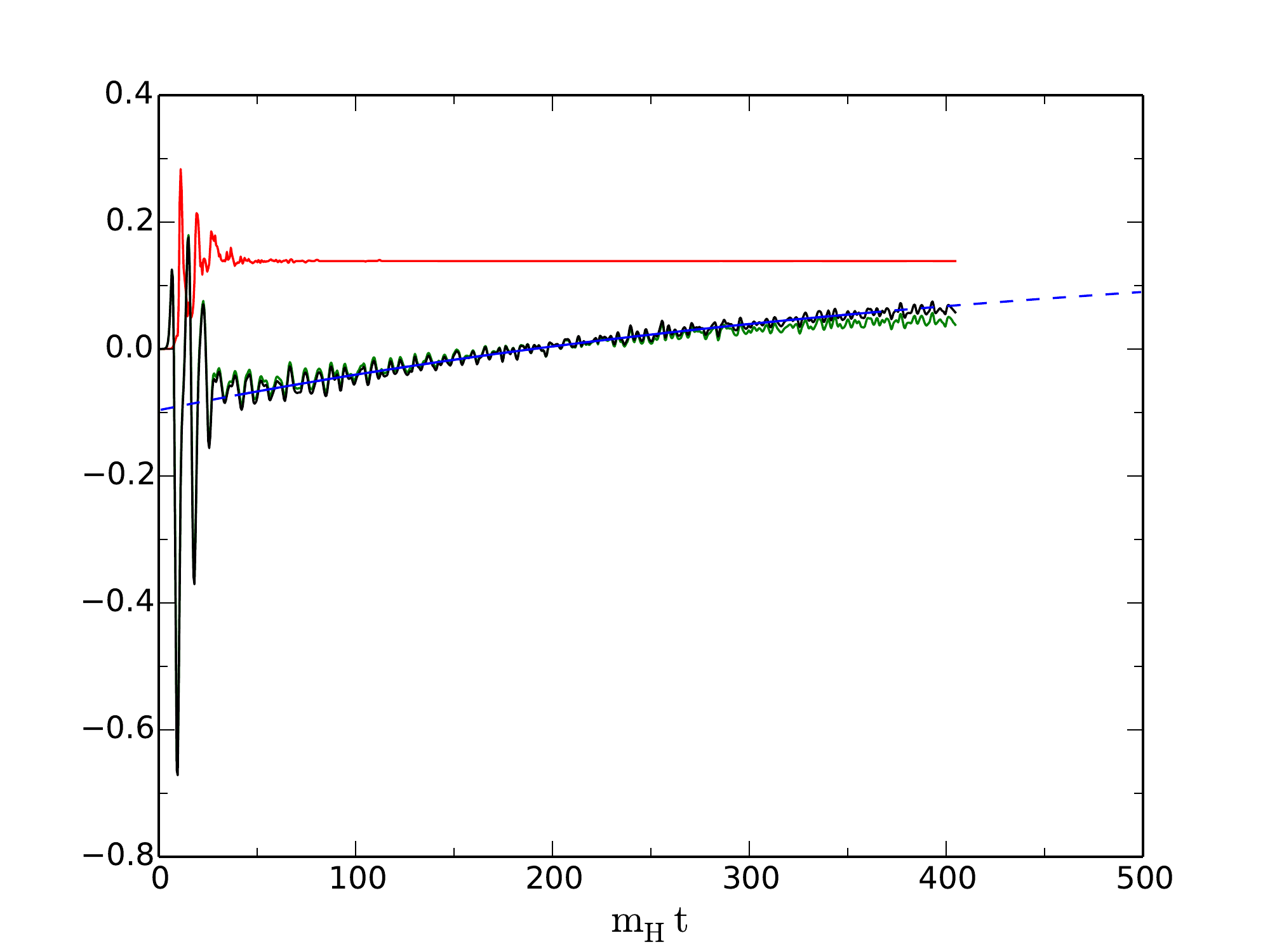} 
\end{tabular}
\caption{The late-time approach of $N_{\rm cs,2}$ to $N_{\rm w}$, for two quench times $m_H\tau_q=0$ and $16$, with $\delta_{\rm cp}=6.83$. }
\label{fig:Longncs2}
\end{figure}

We first display in Fig.~\ref{fig:Ex_averages} a typical set of averaged observables as a function of time, for a particular quench time $m_H\tau_q=16$ and CP-violation $\delta_{\rm cp}=6.83$. We observe the rolling off of the Higgs field and the creation of the baryon asymmetry (here, $N_{\rm w}$ and $N_{\rm cs,2}$) first going up, then sharply down, then up again in a sequence of driven oscillations. We confirm in Fig.~\ref{fig:Longncs2} that for very late times the average Chern-Simons number converges to agree with winding number, and find that the Chern-Simons number fits an exponential curve for the two quench-times in the figure\footnote{The asymptotic value is not exactly $N_{\rm w}$ because of statistical errors. Explicitly requiring that $N_{\rm cs,2}\rightarrow N_{\rm w}$ also produces good fits.}
\ba
\left.N_{\rm cs,2}\right|_{m_H\tau_q=0}\simeq N_{\rm w,\infty}+0.010-0.294\exp(-0.003\,m_Ht),\\
\left.N_{\rm cs,2}\right|_{m_H\tau_q=16}\simeq N_{\rm w,\infty}+0.038-0.273\exp(-0.002\,m_Ht).
\ea
From this fit, one may estimate the time-scale for the convergence to be around $m_Ht\simeq 300-500$. Moreover, we see from Fig.~\ref{fig:Longncs2} that the winding number stays put after $m_Ht=50$, and so in some of our simulations we will only simulate until $m_Ht\simeq 100$ or $400$, and infer the final asymmetry from the value  of $N_{\rm w}$. In this figure, we show the Chern-Simons number computed both by time-integration of (\ref{eq:Ncstime}) and through the spatial definition (\ref{eq:Ncsspace}). A small discrepancy starts to appear around time $m_Ht=300$, due to accumulated errors from the time integration. 

\begin{figure}
\centering
    \includegraphics[width = 0.75\textwidth]{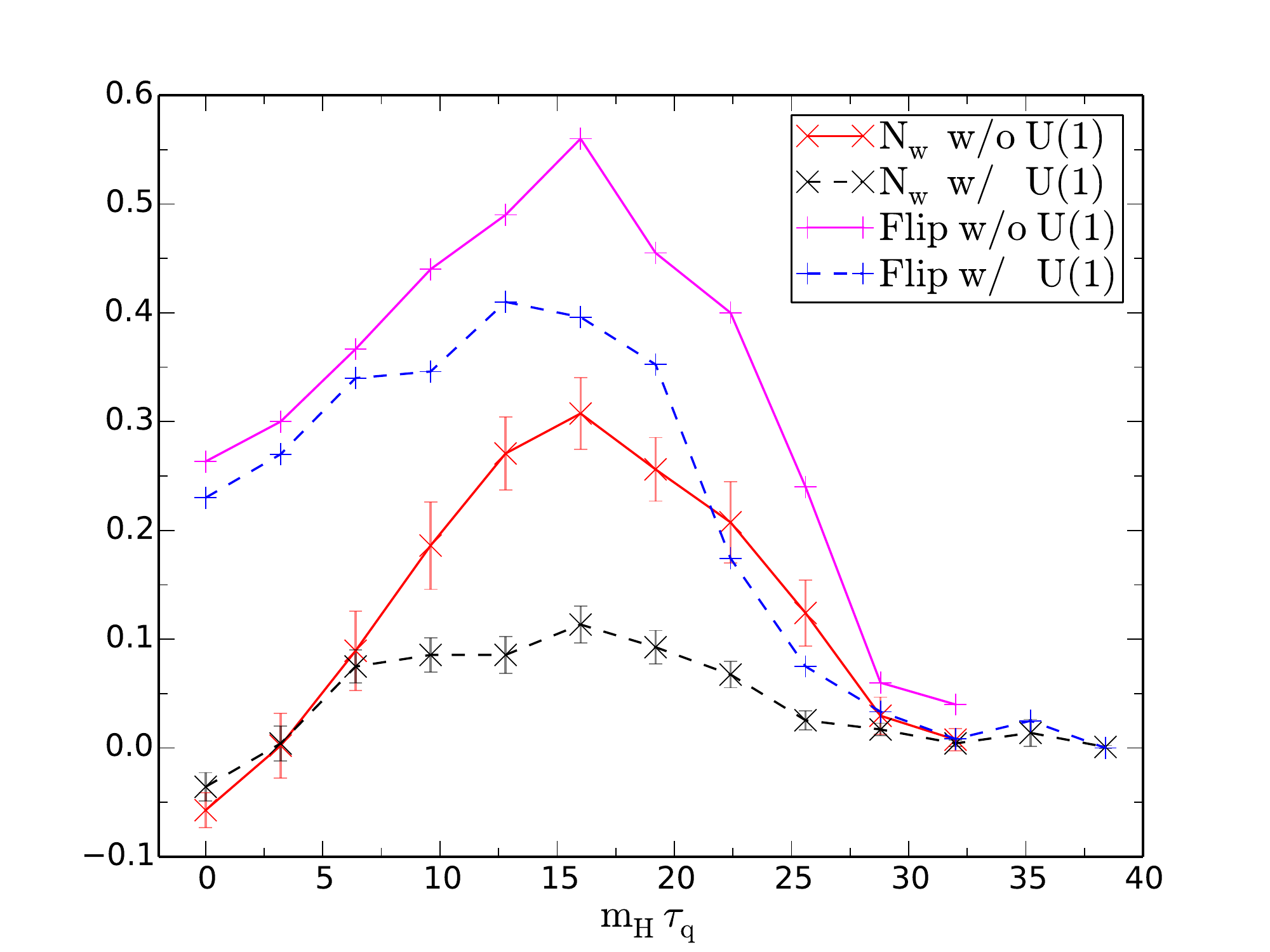}
\caption{The quench time dependence of the final asymmetry, with and without U(1) hypercharge, taking $\delta_{\rm cp}=6.83$.}
 \label{fig:Nw_Adep}
\end{figure}

\subsection{Hypercharge impact on asymmetry}
\label{sec:U1impact}
 
In Fig.~\ref{fig:Nw_Adep}, we show the dependence of the final asymmetry on quench time with the gauge group SU(2) \cite{Jan_quench,CBquench} and with SU(2)$\times$U(1). 
The main features of the quench-time dependence are that the fastest quench ($\tau_q=0$) produces a baryon asymmetry of negative sign, while for intermediate quench times a peak of positive asymmetry appears around $m_H\tau_q\simeq 16$. For slow quenches, $m_H\tau_q>32$, no asymmetry is produced. This is true both with and without the hypercharge field, and in fact for the fastest quenches ($m_H\tau_q<6$), no difference can be seen. The peak at intermediate times is visible, but much less pronounced and flatter, suggesting that the U(1) gauge field has a significant moderating effect on whether CP-violation can bias CP-conjugate pairs of configurations in the ensemble. Numerically, the maximum asymmetry is reduced by a factor of three. The asymmetry again decreases for large quench times, becoming indistinguishable from zero around $m_H\tau_q=25$. 

\begin{figure}
\begin{tabular}{lr}
\hspace{-1cm}
\includegraphics[width = 0.55\textwidth]{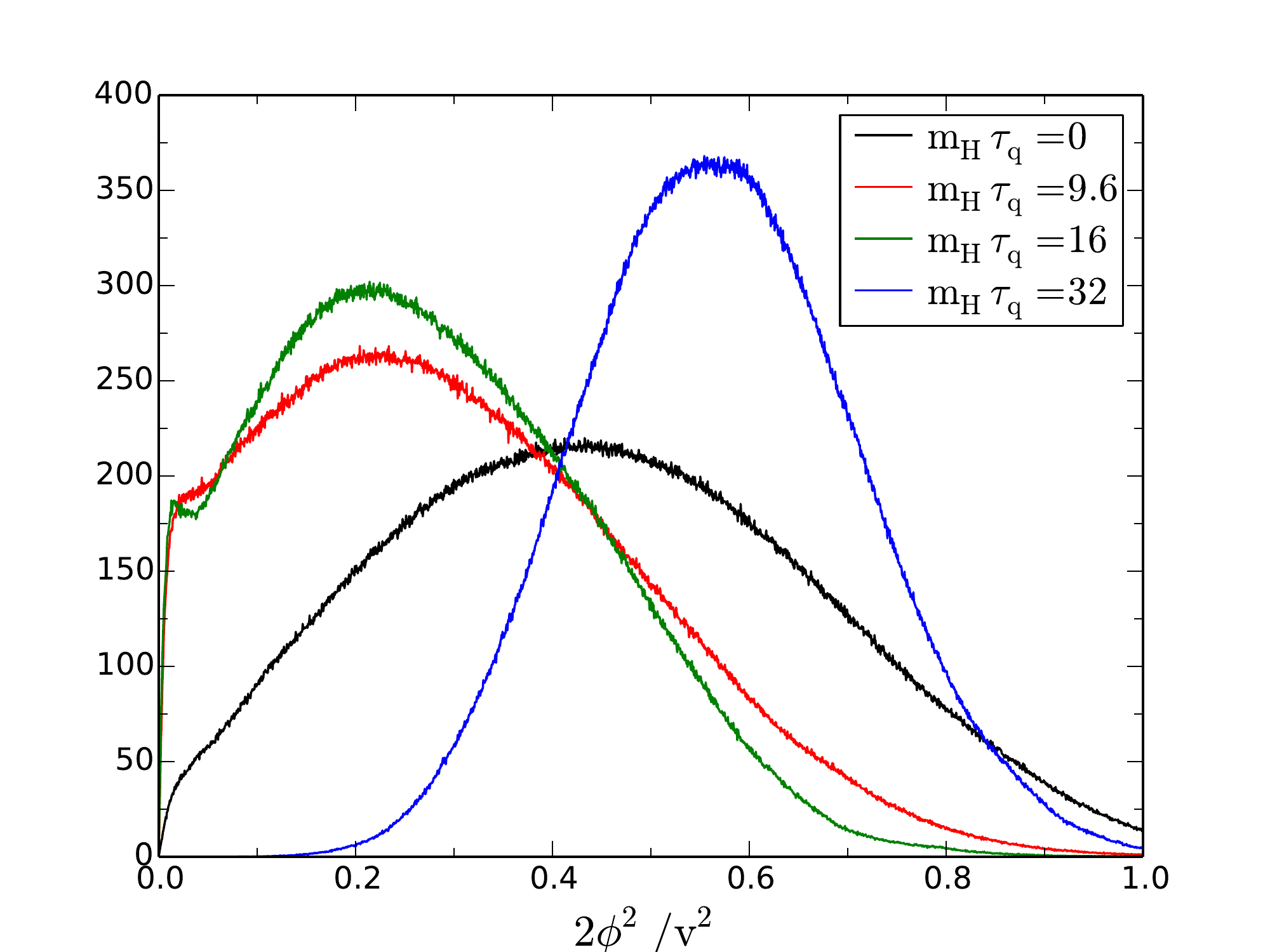} & \hspace{-1.2cm} 
\includegraphics[width = 0.55\textwidth]{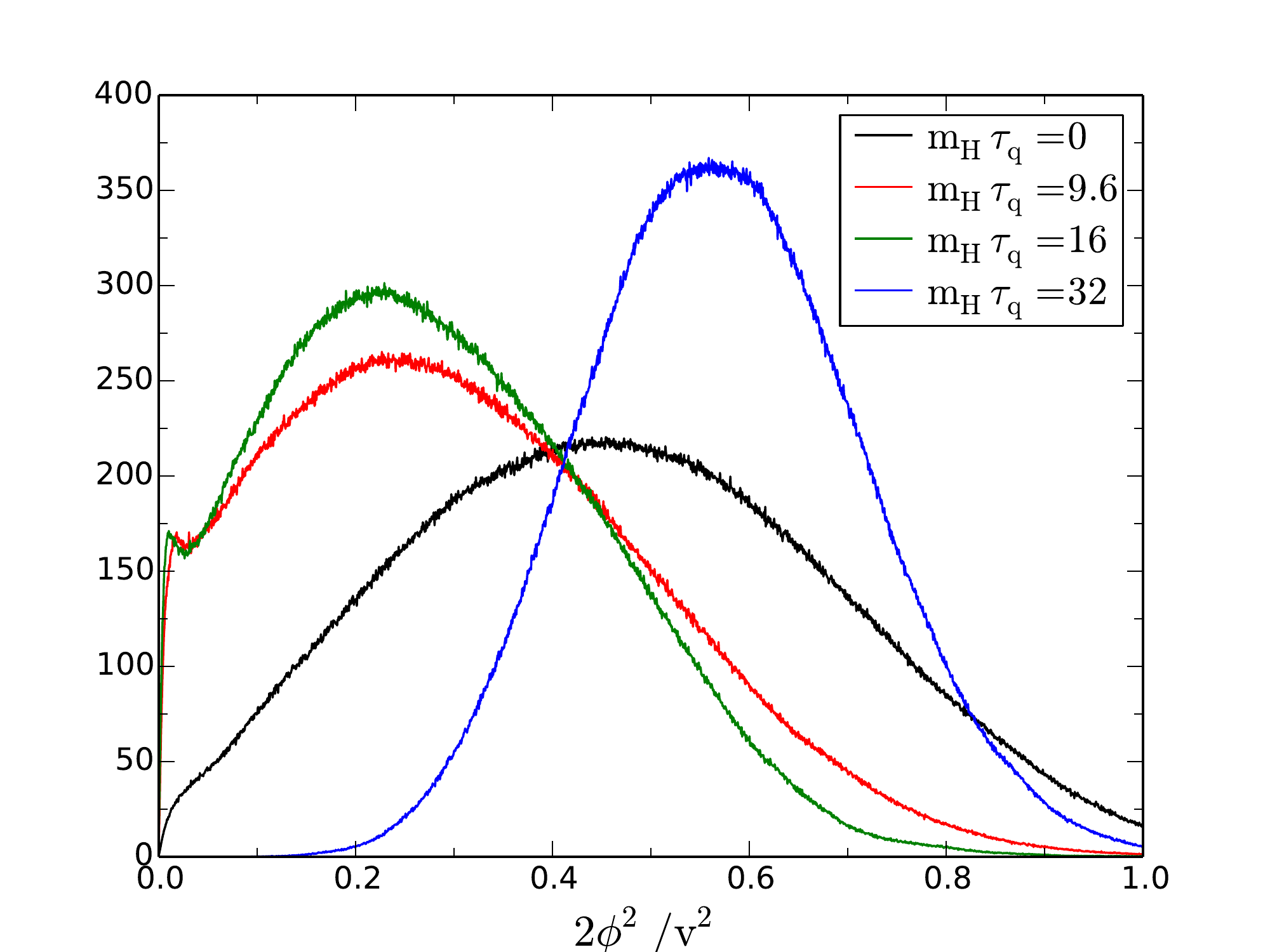} \\ 
\hspace{-1cm}
\includegraphics[width = 0.55\textwidth]{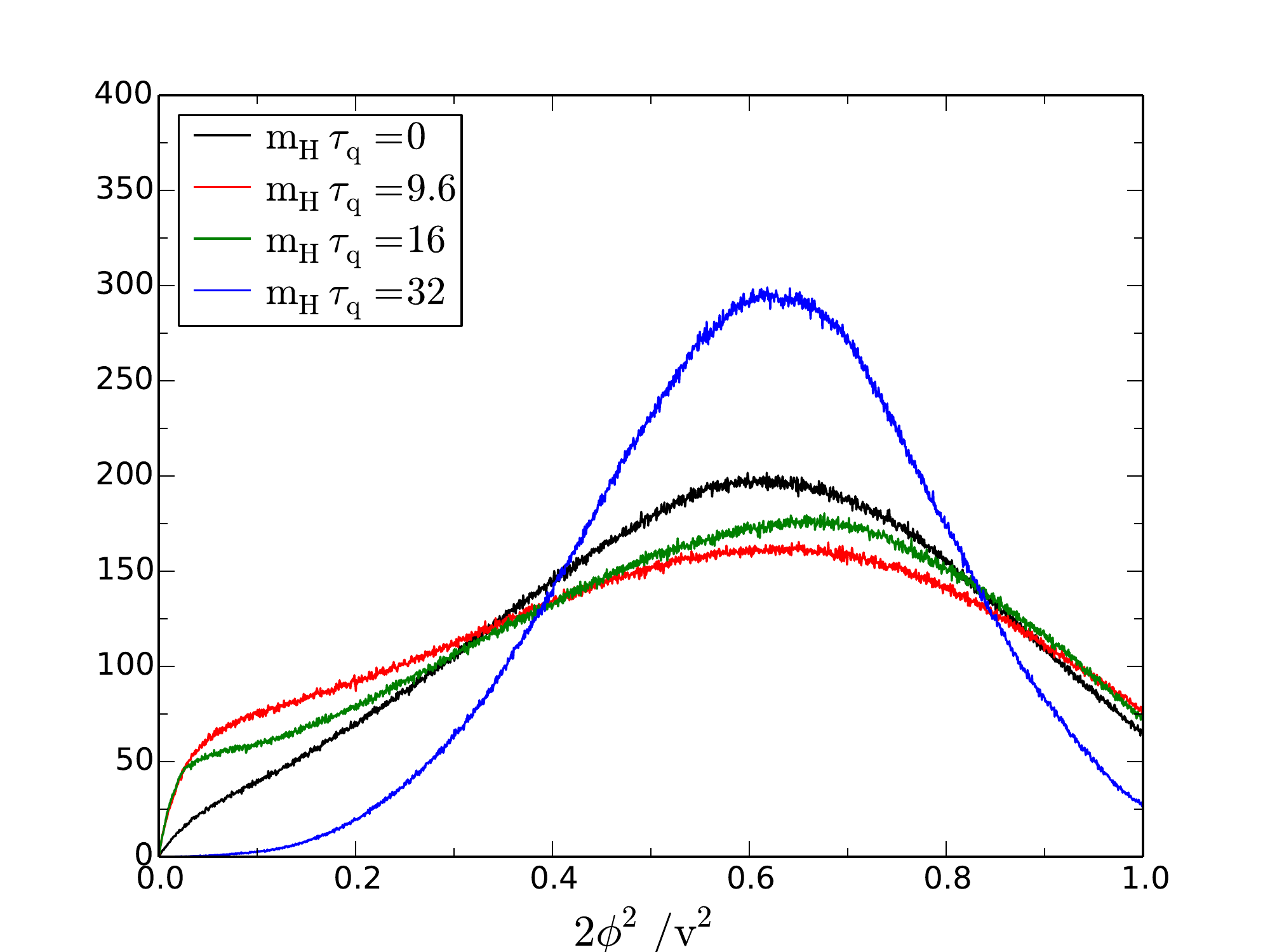} & \hspace{-1.2cm} 
\includegraphics[width = 0.55\textwidth]{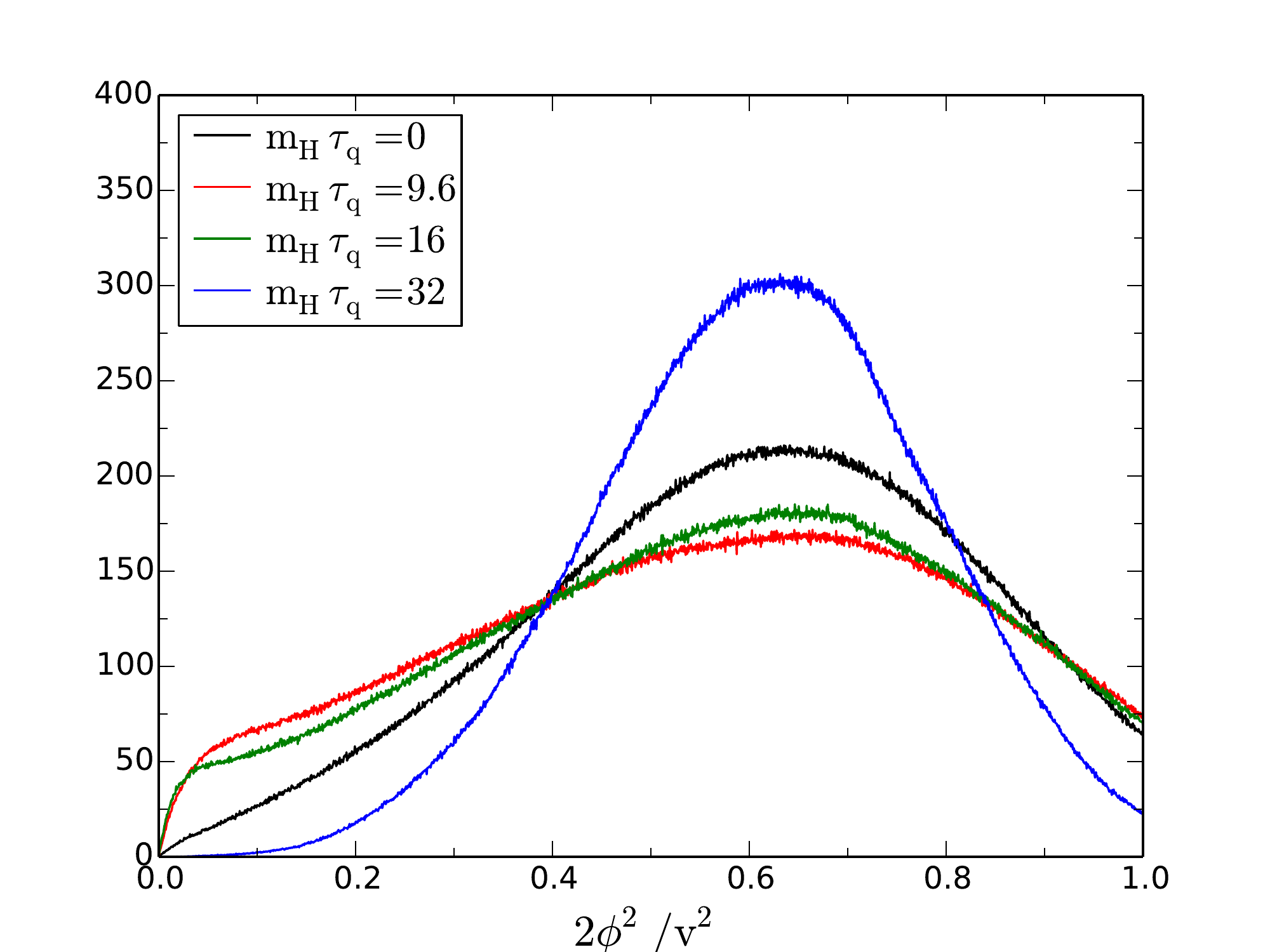}
\end{tabular}
\caption{The distribution of $\phi^2(x)$ over the lattice for different quench times at the first (top line) and second (bottom line) minimum. With U(1) hypercharge (left column) and without (right column). }
 \label{fig:Minima}
\end{figure}

In search for an explanation of this behaviour, we in the same figure (\ref{fig:Nw_Adep}) show the total number of flips (positive and negative) with and without U(1)-fields\footnote{We omit explicit statistical error bars on the ``flip" observable. It is a binary object, whereas jumps are not always integer, nor restricted to unity. We find that errors are comparable to errors on the asymmetry itself.}. We firstly see a clear correlation between the overall number of flips and the asymmetry between positive and negative such flips. The suppression due to the introduction of hypercharge is evident also in the total number of flips, although by a smaller factor. This leads us to consider the number of Higgs-zeroes available for flipping in the first place. Fig.~\ref{fig:Minima} shows histograms of values of $\phi^2$ locally in space for different quench times $m_H\tau_q=0,9.6,16,32$ , with and without U(1) in the simulation. We see that the number of zeros of the Higgs field is strongly correlated with the final asymmetry (this was also observed in \cite{Jan_quench}). But we see very little impact of including the hypercharge U(1). 

The precise mechanism whereby the asymmetry is reduced remains unclear. But we tentatively interpret our data to indicate that the baryogenesis process unfolds in two parts. First the initial roll-off, where the asymmetry is driven negative in the very early stages of the transition. And then a second stage, where the asymmetry is driven up again in consecutive minima of the Higgs field oscillation.  For the fastest quench time, only the first process plays a role, and at that stage the U(1) gauge field has rather little energy associated to it (Fig. \ref{fig:E_average}) and is not large enough to have an impact. For the slower quenches, the second stage is dominant, during a time where the U(1) field is large enough to play a part. Fewer flips occur and a smaller asymmetry is generated.

\begin{figure}
\begin{tabular}{lr}
\hspace{-1cm}
\includegraphics[width = 0.55\textwidth]{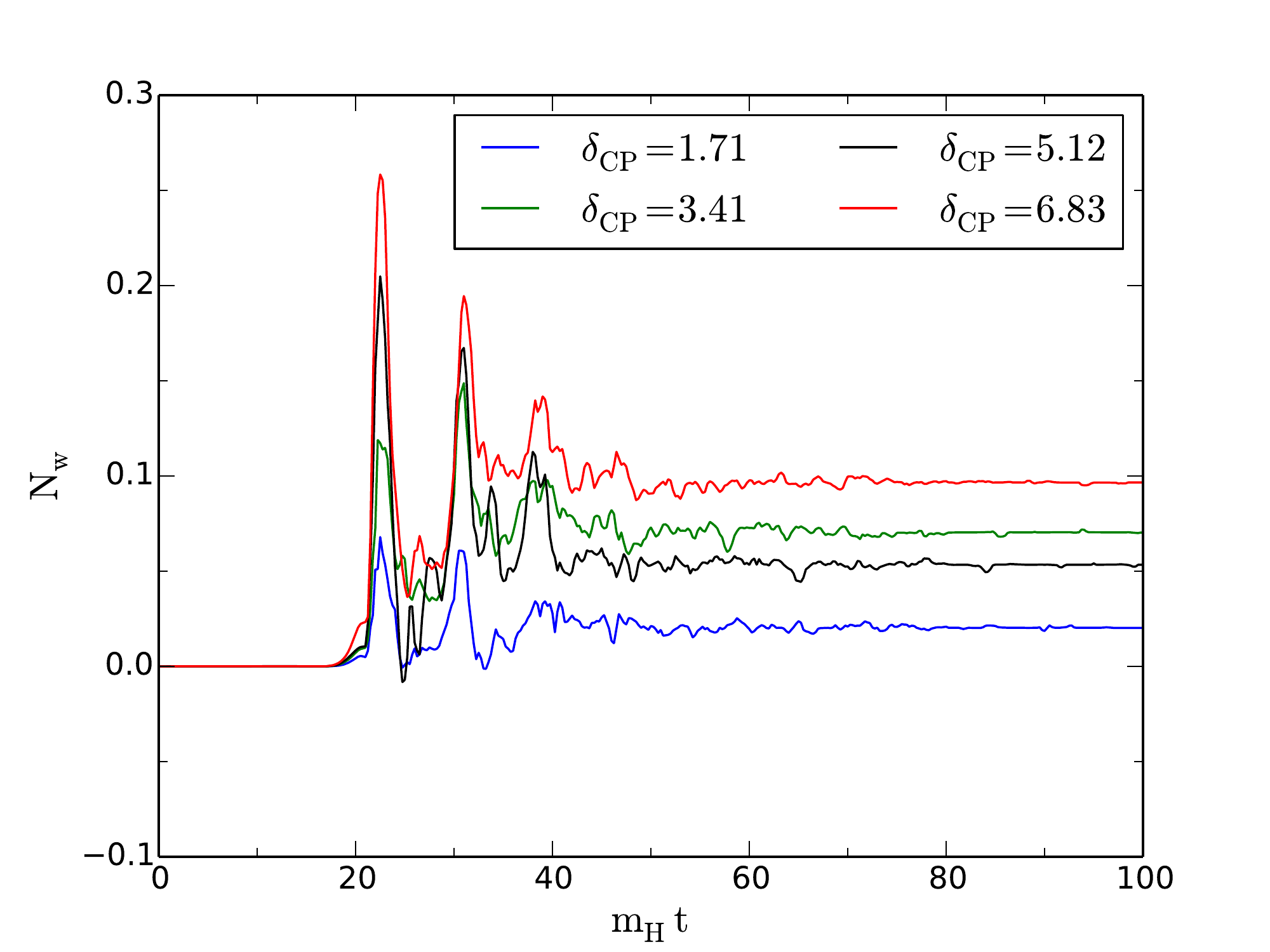} & \hspace{-1.2cm} 
\includegraphics[width = 0.55\textwidth]{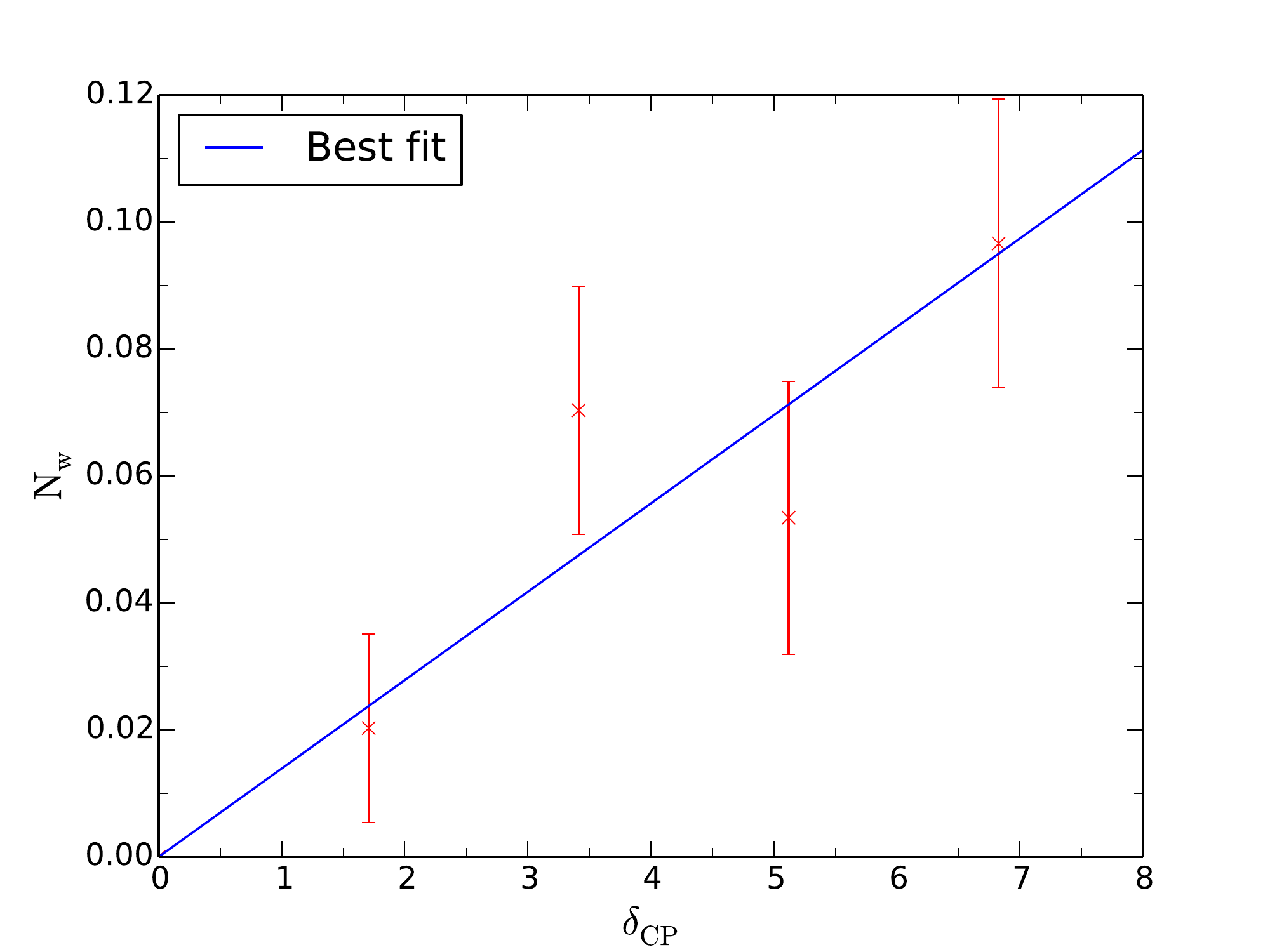} 
\end{tabular}
\caption{The dependence of the ensemble-average of the Higgs winding, $N_{\rm w}$ on the CP violation parameter $\delta_{\rm cp}$. The left figure shows the time dependence of $ N_{\rm w}$ for different values of $\delta_{\rm cp}$, and the figure on the right gives a fit of the final value as a function of $\delta_{\rm cp}$. }
\label{figNw_deltaCP}
\end{figure}
Noting from Fig. \ref{fig:Nw_Adep} that the maximum asymmetry occurs with $m_H\tau_q\simeq 16$, we now wish to see how the CP violation affects the final Chern-Simons number. To this end we simulated an ensemble of runs with quench time $m_H\tau_q=16$ and show in Fig. \ref{figNw_deltaCP} (left) the time dependence of the ensemble average of $N_{\rm w}$ for four different values of $\delta_{\rm cp}$. Then in Fig. \ref{figNw_deltaCP} (right) we fit the final value of $N_{\rm w}$ vs $\delta_{\rm cp}$, and see that the final Higgs winding, or equivalently the final Chern-Simons number, is linear in $\delta_{\rm cp}$
\ba
N_{\rm w}(t\to\infty)=1.39\times 10^{-2}\delta_{\rm cp}.
\ea
Note that we us a one-parameter fit, since the intercept at $\delta_{\rm cp}=0$ vanishes by virtue of the CP-symmetric ensemble.

\subsection{Helical magnetic field and baryogenesis}
\label{sec:chiral}

\begin{figure}
\begin{tabular}{lr}
\hspace{-1cm}
\includegraphics[width = 0.55\textwidth]{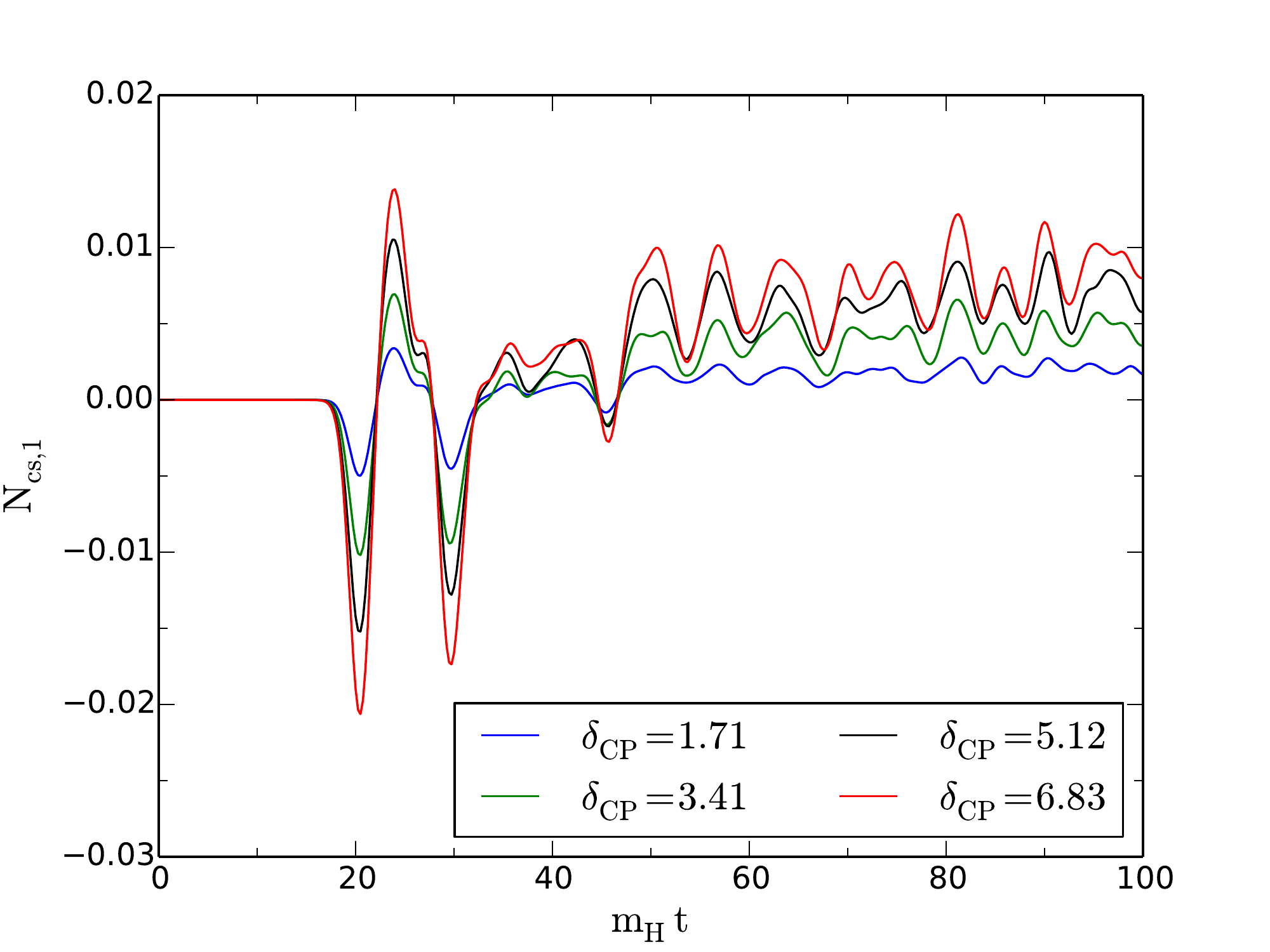} & \hspace{-1.2cm} 
\includegraphics[width = 0.55\textwidth]{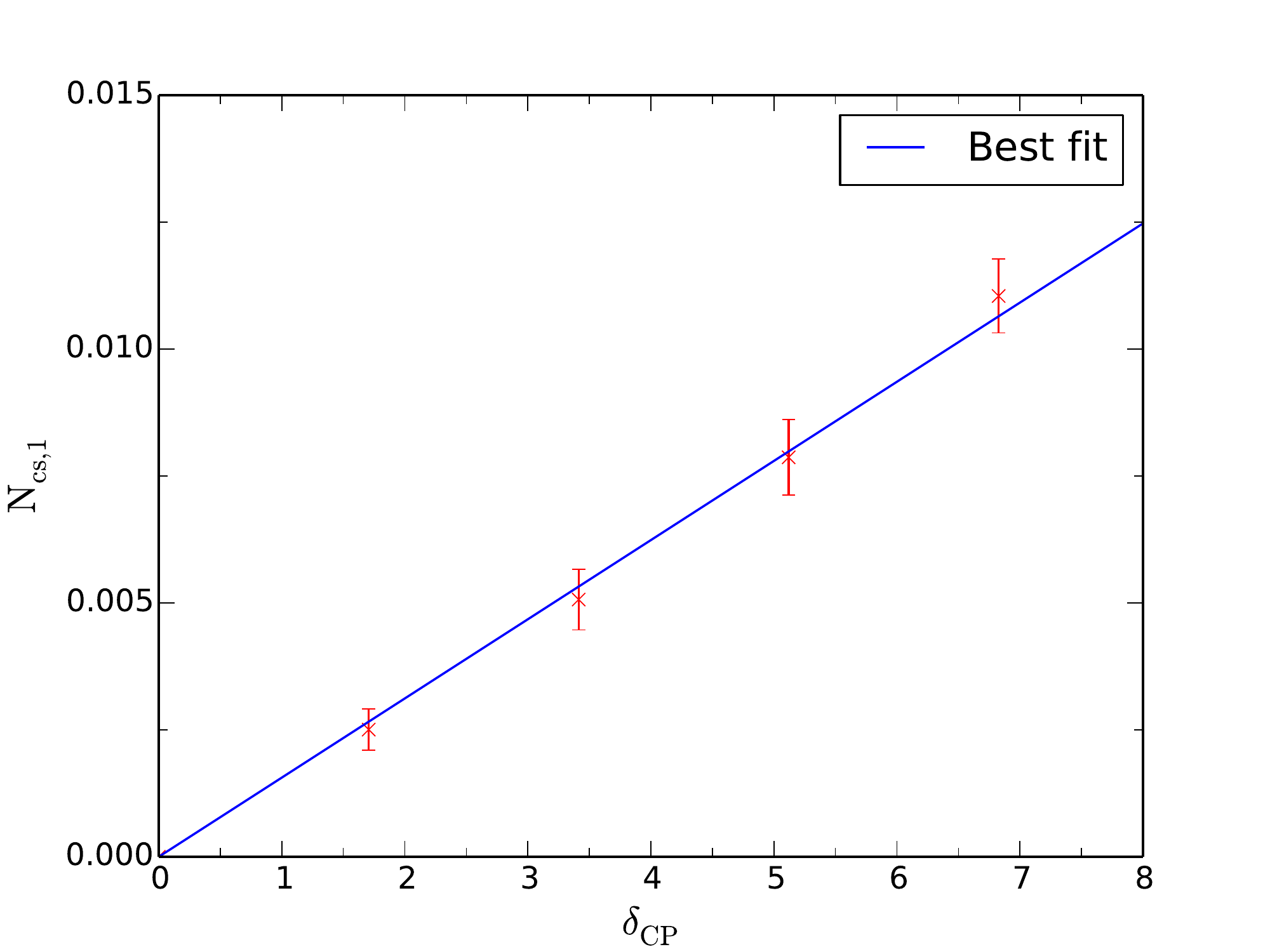} 
\end{tabular}
\caption{The dependence on CP-violation strength $\delta_{\rm cp}$ of the hypercharge Chern-Simons number, with $m_H\tau_q=16$. }
 \label{fig:CPdepncs1}
\end{figure}
We next turn to the other CP-odd quantities $N_{\rm cs,1}$ and $N_{\rm h}$. In the absence of CP-violation, these are also forced by our CP-symmetric ensemble to average to zero. Once $\delta_{\rm cp}\neq 0$, they are allowed to stray from zero average, but whereas our CP-violating term explicitly biases $N_{\rm cs,2}$ (the ``primary" biased observable), it is not obvious in what way this influences what we will term the ``secondary" biased observables. $N_{\rm w}$ is also secondary in this terminology, and we have seen that it is pushed around by a violently oscillating $N_{\rm cs,2}$ to eventually settle and determine the final baryon asymmetry. In a similar way, we would expect $N_{\rm cs,1}$ and $N_{\rm h}$ to be biased by evolving in the background of non-zero $N_{\rm cs,2}$.

In Fig.~\ref{fig:CPdepncs1} we show the time evolution of $N_{\rm cs,1}$ (left) and the final asymmetry (right) for different values of $\delta_{\rm cp}$, for quench time $m_H\tau_q=16$ . We see a very clear linear dependence on the strength of CP-violation, and we can fit this dependence to find 
\begin{eqnarray}
\langle N_{\rm cs,1}\rangle\simeq1.56\times 10^{-3} \delta_{\rm cp},
\end{eqnarray}
for this quench time. The time evolution is particularly striking, in that in the early stages curves of different $\delta_{\rm cp}$ oscillate completely coherently, even crossing zero at the same time. We also note that because the observables take on both positive and negative values in quite a complicated evolution, the chances of finding a well-motivated effective chemical potential function to reproduce this behaviour in linear response is remote.

\begin{figure}
\begin{tabular}{lr}
\hspace{-1cm}
\includegraphics[width = 0.55\textwidth]{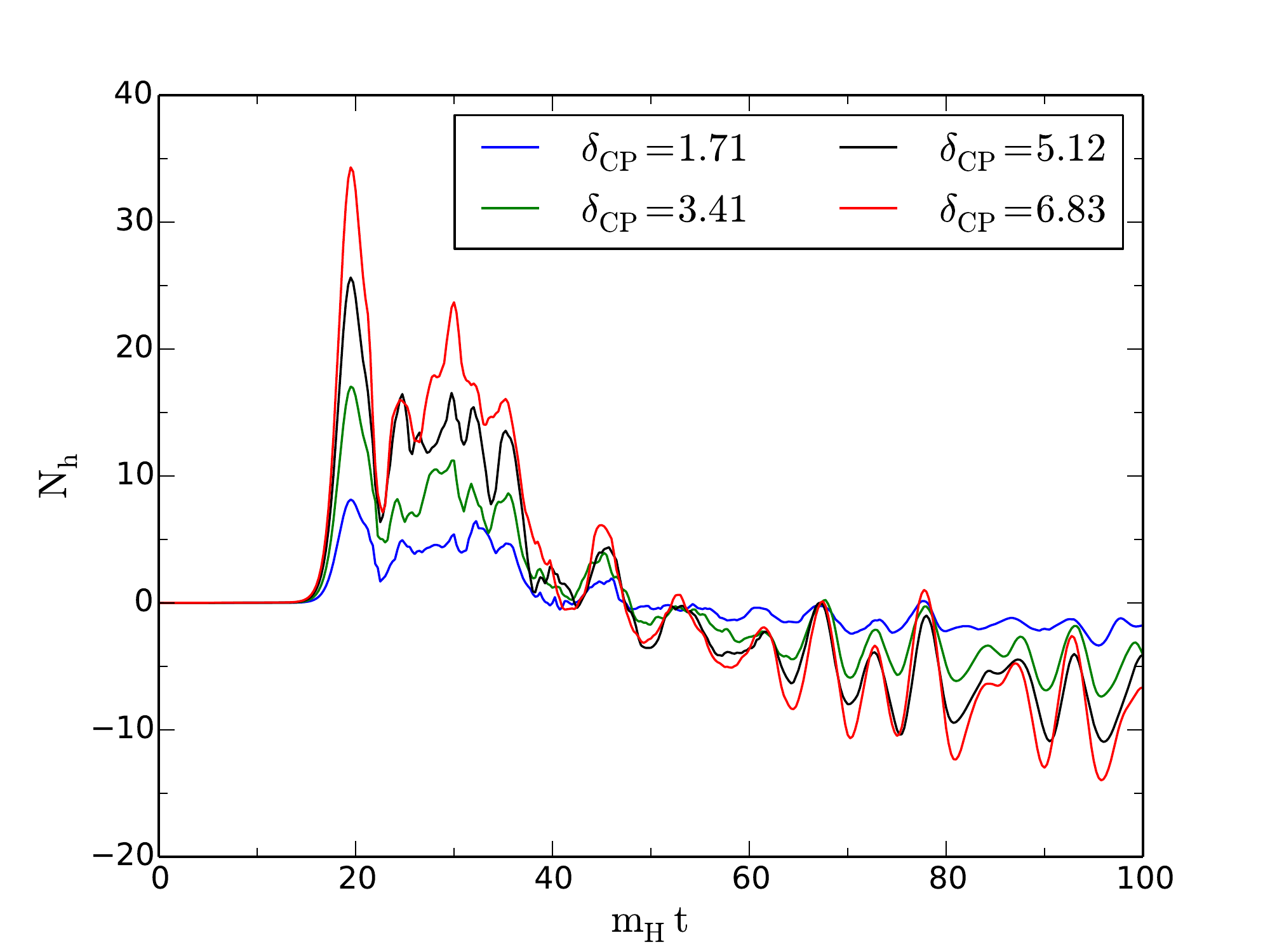} & \hspace{-1.2cm} 
\includegraphics[width = 0.55\textwidth]{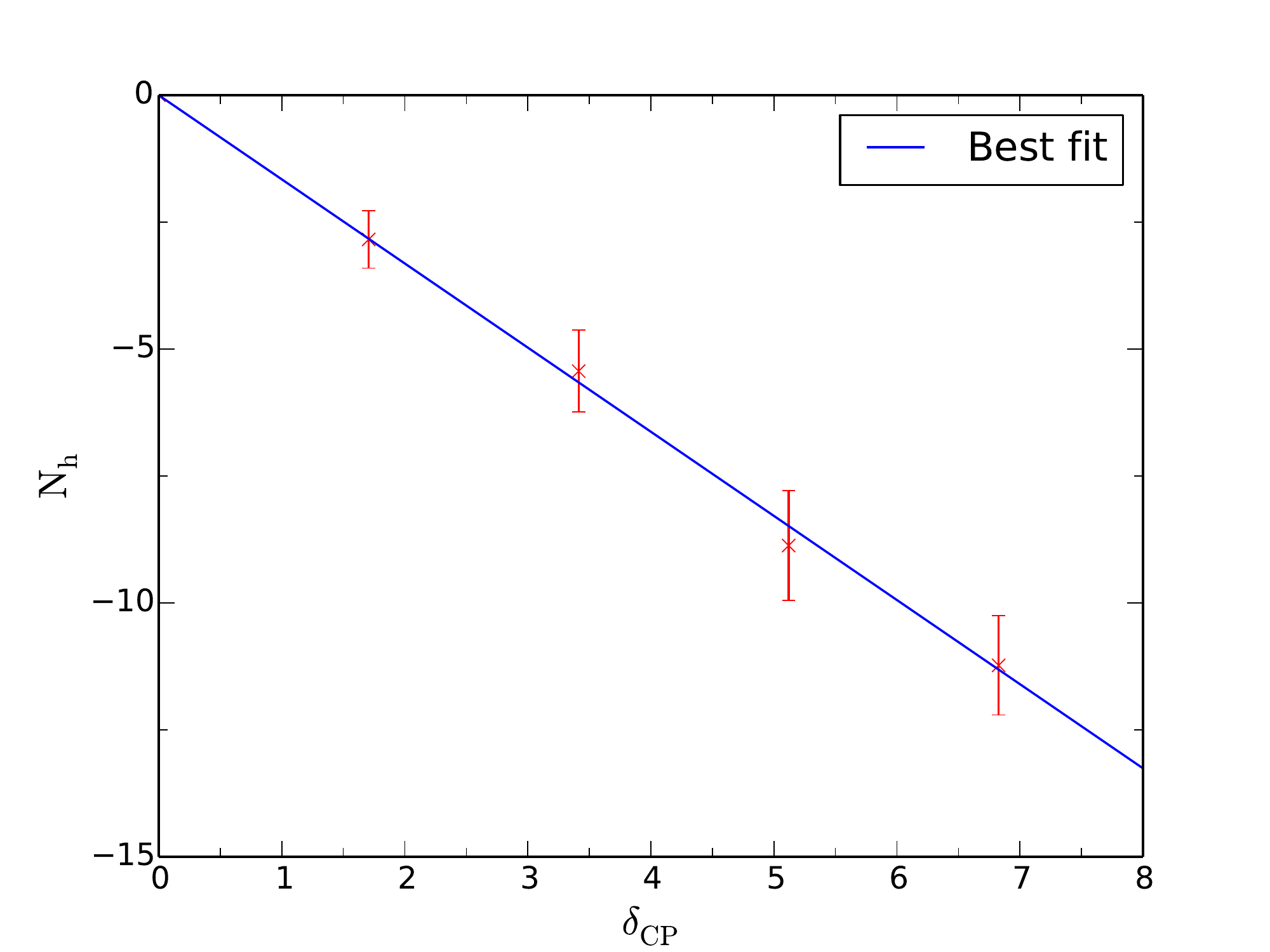} 
\end{tabular}
\caption{The dependence on CP-violation strength $\delta_{\rm cp}$ of the magnetic helicity, with $m_H\tau_q=16$.   }
 \label{fig:CPdepmh}
\end{figure}
In Fig.~\ref{fig:CPdepmh} we show the magnetic helicity, for the same simulations. Again, we see that the dependence on $\delta_{\rm cp}$ is close to linear, and we find
\begin{eqnarray}
\langle N_{\rm h}\rangle\simeq -1.66 \,\delta_{\rm cp},
\end{eqnarray}
for this quench time of $m_H\tau_q=16$. 

\begin{figure}
\begin{tabular}{lr}
\hspace{-1cm}
\includegraphics[width = 0.55\textwidth]{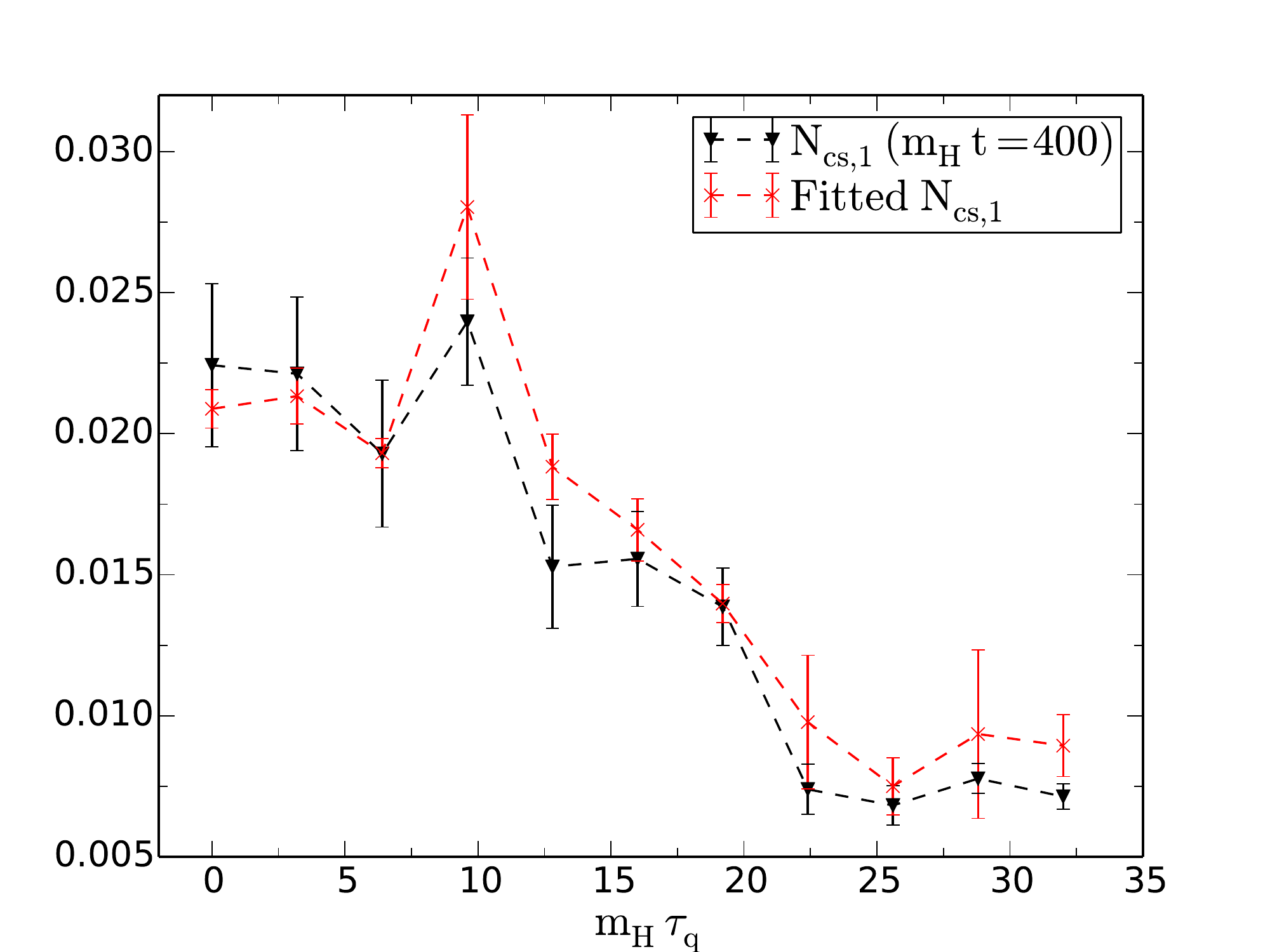} & \hspace{-1.2cm} 
\includegraphics[width = 0.55\textwidth]{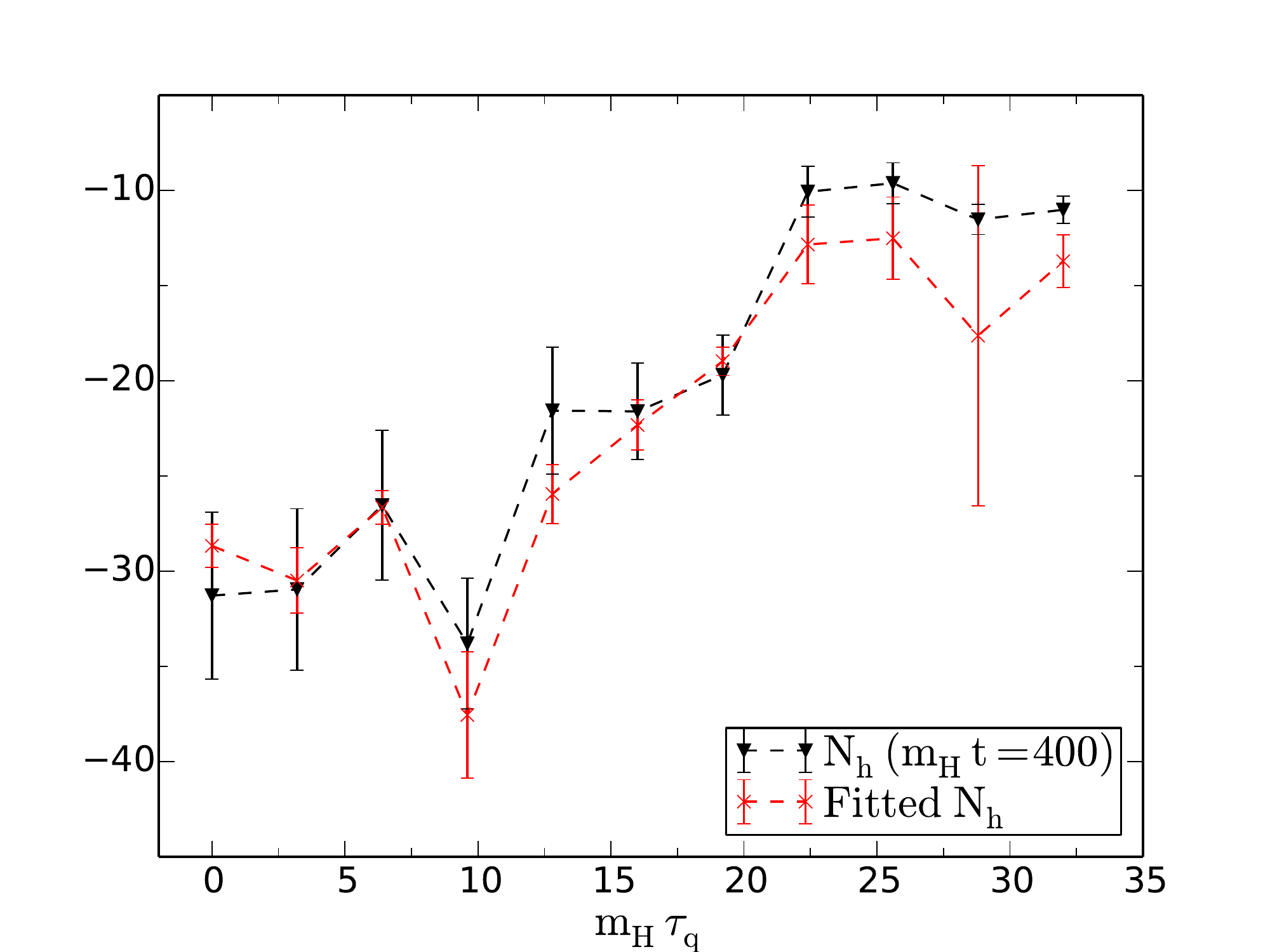} 
\end{tabular}
\caption{The late-time quench time dependence of $N_{\rm cs,1}$ (left) and $N_h$ (right), with $\delta_{\rm cp}=6.83$. }
 \label{fig:Odd_quench}
\end{figure}
We show in Fig.~\ref{fig:Odd_quench} the average hypercharge Chern-Simons number (left) and magnetic helicity (right) as a function of quench time. 
We have run to time $m_Ht=400$, and extrapolated to an asymptotic value as for $N_{\rm cs,2}$ in Fig.~\ref{fig:Longncs2}. We see that for both observables, the dependence on quench time is monotonic. Largest asymmetry for fast quenches, small for slow quenches. The quenches are not so slow that the magnetic helicity disappears altogether. The final temperature is still about 50 GeV and some of this energy goes into the magnetic field, also generating helicity. 

In \cite{tanmay1}, the prediction is that $\langle N_h\rangle\simeq -300 \langle N_{\rm cs,2}\rangle$ for the decay of Sphalerons. By order of magnitude, this is correct for $m_H\tau_q=0$, whereas for all the other quench times, the sign is opposite. The crucial point is that the quench time dependence of the magnetic helicity, Fig. \ref{fig:Odd_quench}, is qualitatively different from the dependence of $N_{\rm cs,2}$, Fig. \ref{fig:Nw_Adep}. We conclude that the non-zero signal in the secondary observables indicate the presence of CP-violation (in this case, that $N_{\rm cs,2}$ is non-zero), as well as the possibility of baryogenesis, but they are not a reliable proxy to infer the magnitude of the baryon asymmetry.

\subsection{Sphaleron unravelling}
\label{sec:sphaleron}

\begin{figure}
\centering
    \includegraphics[width = 0.65\textwidth]{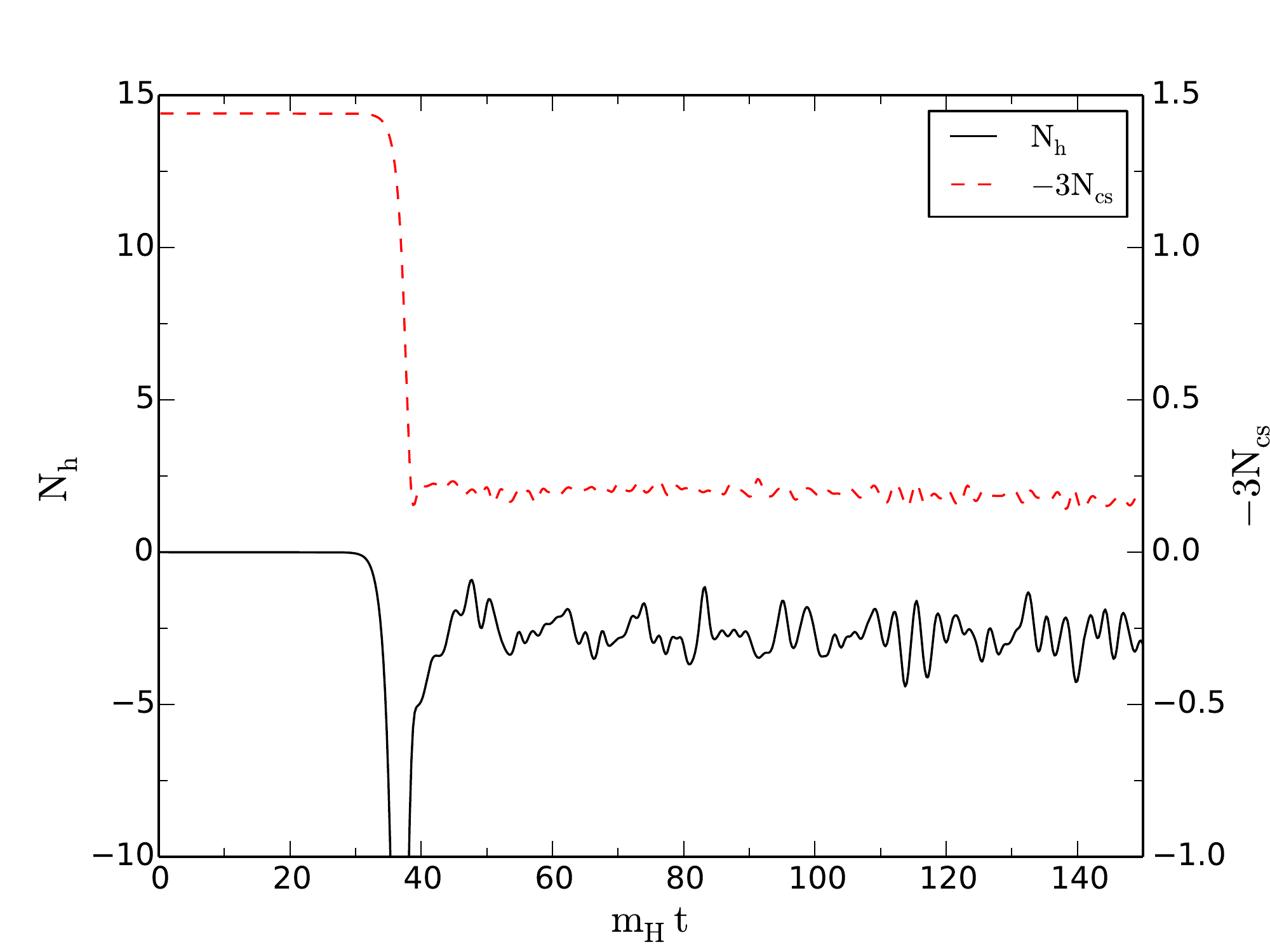}
\caption{The Chern-Simons number and helicity as a Sphaleron unravels.}
 \label{fig:sphaleron}
\end{figure}

In this final section, we offer an example of the different mechanisms at play in \cite{tanmay1, tanmay2} and the present work, to further emphasize that there is no direct proportionality between magnetic helicity and Chern-Simons number. In Fig.~\ref{fig:sphaleron}, we show the SU(2) Chern-Simons number and the helicity during a sphaleron decay. This is a direct reproduction of the simulation in \cite{tanmay2}, where we set up the initial conditions close to the sphaleron solution \footnote{If one uses the actual sphaleron solution it will not decay as it is a static, albeit unstable, solution to the equations of motion.}. As one would expect, the change in Chern-Simons number for the sphaleron decay is one-half, given that the sphaleron is the configuration that is half way between integer Chern-Simons number vacua\footnote{In the plot we multiply $N_{\rm cs}$ by $-3$ to match the convention used in  \cite{tanmay2}}. The unravelling of the sphaleron then leads to the formation of helical magnetic fields, with $N_h\sim 2-3$.

\begin{figure}
\centering
    \includegraphics[width = 0.65\textwidth]{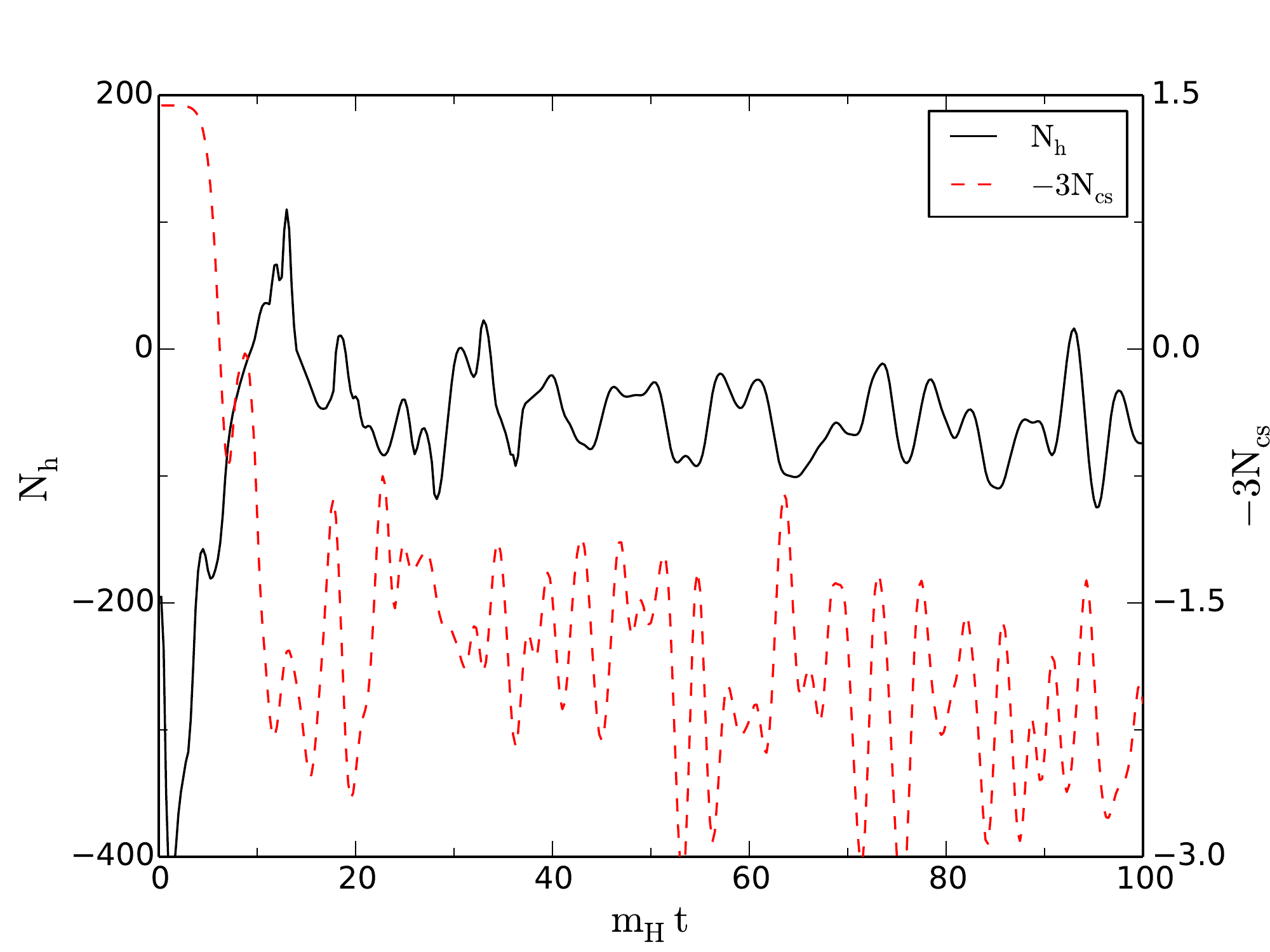}
\caption{The Chern-Simons number and helicity during a tachyonic roll-off.}
 \label{fig:nosphaleron}
\end{figure}
In Fig.~\ref{fig:nosphaleron}, we show the same observables, but during the tachyonic roll-off of a single configuration. We see that the structure leads to a rather different relationship between $N_h$ and $N_{\rm cs}$, the signal has a large stochastic component, and although the change in Chern-Simons number is one, the helicity fluctuates around -100.

Sphalerons play a key role in Hot Electroweak Baryogenesis as they represent the dominant trajectory for Chern-Simons number change in thermal equilibrium. That particular trajectory involves Chern-Simons number and winding changing at the same time, because that is energetically favourable. Cold Electroweak Baryogenesis is violently out of equilibrium, and Chern-Simons number may change along many different trajectories, without reference to winding number. As a result, Sphaleron configurations play no role for the baryogenesis process, except at very late times when the system is close to equilibrium.

\section{Conclusion}
\label{sec:conc}
We have explored the physics of baryogenesis and helical magnetogenesis using the first simulations that include both CP-violation, and the full Bosonic sector of the Standard Model. The scenario under investigation is that of Cold Electroweak Baryogenesis, whereby the fields are taken out of equilibrium due to an exponential growth of the IR modes, rather than a thermal phase transition. An important property of such a scenario is the quench time, namely how long it takes for the Higgs potential to evolve from a symmetry-restored shape, to the symmetry-broken form. Our simulations have confirmed a similar result found with the absence of U(1) hypercharge \cite{Jan_quench,CBquench} that there is a non-zero value of the quench time that maximizes the creation of Higgs winding, and so also Chern-Simons number. One thing that turned out to be of practical importance was that while $N_{\rm cs,2}$ had a relaxation timescale of $\sim 300\,m_H^{-1}$, the winding number became fixed at around $50\,m_H^{-1}$, allowing us to infer the final Chern-Simons number from the winding number. 

The effect of the introduction of U(1) hypercharge is to reduce the final asymmetry for all values of quench-time, but there remains a peak value for quench times of around $m_H\tau_q\simeq16$. This peak, however, is reduced by a factor of about three, and also flattened. One thing that the U(1) hypercharge does not affect though is the number of zeros of the Higgs field at minima of its mean value.

As these simulations also include the hypercharge we are able to measure the magnetic field that is formed during the spinodal instability, and subsequent evolution. In particular, we focussed on the helical component of the magnetic field, as this has been proposed as a measure of baryogenesis \cite{tanmay1, tanmay2}. In examining the relationship of our observables on the CP-violation and quench time we discovered that, for a fixed quench time, the Chern-Simons number and the magnetic helicity are proportional to $\delta_{\rm cp}$. However, we found that the non-trivial dependence on quench time for these observables meant that the Chern-Simons number could not be assumed to simply follow the final value of helicity. This was further backed up by examining the case of a single sphaleron, compared to a tachyonic falling-off of the Higgs field, finding that helicity is not directly proportional to the Chern-Simons number. The consequence of these results is that an observation of magnetic helicity cannot be used as a proxy for baryogenesis.

On the other hand, we also found that all CP-odd observables were not only non-zero, but that the asymmetries were all proportional to $\delta_{\rm cp}$. This is not surprising for $N_{\rm cs,2}$, which is explicitly (primary) biased by the specific CP-violating term used here. But that the same proportionality applies to the secondary biased $N_{\rm cs,1}$ and $N_h$ shows that they faithfully pick up the CP-breaking through the amplitude of the non-zero $N_{\rm cs,2}$. Conversely, one may imagine a CP-violating theory, where $N_{\rm cs,2}$ is a secondary rather than primary biased observable. One may then expect a baryon asymmetry to be generated there as well. This is currently under investigation \cite{inflaton_us}.

\vspace{0.2cm}

\noindent
{\bf Acknowledgments:}  AT and ZGM are supported by a  UiS-ToppForsk grant from the University of Stavanger. PS acknowledges support by STFC under grant ST/L000393/1. The numerical work was performed on the Abel Cluster, owned by the University of Oslo and the Norwegian metacenter for High Performance Computing (NOTUR), and operated by the Department for Research Computing at USIT, the University of Oslo IT-department.

\appendix

\section{Lattice implementation}
\label{sec:lattice}

The classical action on the lattice can be written as  $S= d^4x \sum_{x,t} {\mathcal L}$, with the lagrangian
\begin{eqnarray}
\label{equ:lagrangian}
{\mathcal L} &=& \sum_i \frac{2}{(g'dt\, dx_i)^2} \left(2-V_{0i} - V_{i0}\right) 
 -\sum_{ij}\frac{1}{(g'dx_idx_j)^2} \left(2-V_{ij} - V_{ji}\right) 
\nonumber \\ & &
 + \sum_i \frac{4}{(gdtdx_i)^2} \left( 1- \frac{1}{2} {\rm Tr}\, U_{0i}\right) 
 - \sum_{ij} \frac{2}{(gdx_idx_j)^2} \left( 1- \frac{1}{2} {\rm Tr}\, U_{ij}\right) 
\nonumber \\ & &
 +\frac{1}{2} {\rm Tr} \left[ \dot \Phi^\dagger \dot \Phi \right]
 -\frac{1}{2} \sum_i {\rm Tr} \left[ (D_i \Phi)^\dagger (D_i \Phi) \right]
 -\frac{m_H^2}{2v^2}  \left(  \frac{1}{2} {\rm Tr} [ \Phi^\dagger\Phi ] - \frac{v^2}{2}c_Q \right)^2
\nonumber \\ & &
+\Delta {\mathcal L_2},
\end{eqnarray}
where $c_Q=\left(2\frac{t}{\tau_Q}-1\right)$ when $t<\tau_Q$; $c_Q=1$ when $t\geq \tau_Q$ and $\Delta {\mathcal L_2}$ is the CP-violating term, whose expression will be discussed in section \ref{sec:l2}. The subscript ``0" refers to the time-like direction, and subscripts $i$, $j$, $k$ to space-like directions. We use the short-hand for the Higgs momentum $\dot\Phi=[\Phi(x+0)-\Phi(x)]/dt$.
In the expression above, we have used the plaquette
\begin{eqnarray}
& &U_{\mu \nu}(x)  =   U_\mu(x)U_\nu(x+\mu)U^\dagger_\mu(x+\nu)U^\dagger_\nu(x)
,\nonumber \\
& &V_{\mu \nu}(x)  =   V_\mu(x)V_\nu(x+\mu)V^\dagger_\mu(x+\nu)V^\dagger_\nu(x)
,\end{eqnarray}
and the gauge covariant derivative
\begin{align}
D_\mu \Phi(x) = \frac{U_\mu(x)\Phi(x+\mu) Z_\mu(x) -\Phi(x)}{dx_\mu}.
\end{align}
By interpreting gauge links as exponential functions of gauge fields,
\begin{eqnarray}
& & U_\mu(x) \sim \exp \left(-\frac{i\sigma^a}{2}gdx_\mu W^a_\mu(x) \right),
\quad
V_\mu(x) \sim \exp \left(\frac{i}{2}g'dx_\mu B_\mu(x)\right),
\nonumber\\
& & Z_\mu(x)
=\left( \begin{array}{cc} V^*_{\mu}(x) & 0  \\  0 & V_{\mu}(x) \end{array}\right)
 \sim \exp \left(-\frac{i\sigma^3}{2}g'dx_\mu B_\mu(x)\right),
\end{eqnarray}
we can connect to the continuum lagrangian in the leading order of the perturbation of (\ref{equ:lagrangian})

We further adopt the temporal gauge fixing, so that $U_0=1$ and $V_0=1$, and define electric fields on the lattice as,
\begin{eqnarray}
E_i^a(x)=-\frac{{\rm Tr} \left[ i\sigma^a U_i(x+0) U_i^\dagger(x) \right]}{gdtdx_i}
,\quad
E_i(x)=-\frac{V_{i}(x+0)V_i^\dagger(x) - V_{i}(x)V_i^\dagger(x+0)}{ig'dtdx_i}.\nonumber\\
\end{eqnarray}

The equations of motion can then be derived from the Euler-Lagrangian equation. For the Higgs field:
\begin{align}
\frac{\Phi(x+0)-2\Phi(x)+\Phi(x-0)}{dt^2}& =
\sum_i \frac{U_i(x)\Phi(x+i) Z_i(x) -2\Phi(x) + U_i^\dagger(x-i) \Phi(x-i) Z_i^\dagger(x-i)}{d^2x_i}
\nonumber \\
&- \frac{m_H^2}{v^2} \left(\frac{1}{2}{\rm Tr}[\Phi^\dagger(x) \Phi(x)] - \frac{v^2}{2}c_Q\right) \Phi(x)
 +\sum_\beta\frac{1}{2}\frac{\partial \Delta {\mathcal L_2}}{ \partial \Phi_\beta}k_\beta,
\end{align}
where $k_\beta$ is one $2\times 2$ matrix in $(1,i\sigma^1,i\sigma^2,i\sigma^3)$, so that $\Phi=\sum_\beta\Phi_\beta k_\beta$, and
\begin{eqnarray}
\Phi(x+0) = \Phi(x)+dt\dot\Phi(x).
\end{eqnarray}

For the U(1) gauge field, we find:
\begin{eqnarray}
\hspace{-1cm}
\frac{E_{i}(x)-E_{i}(x-0)}{dt} =
- \sum_j\frac{i}{g'dx_id^2x_j}\left[V_{ij}(x)-V_{ji}(x)+V_{ji}(x-j)-V_{ij}(x-j)\right]- j^{i}_{\Phi} , 
\end{eqnarray}
with
\begin{eqnarray}
V_i(x+0) = \left[\sqrt{1-\left(\frac{g'}{2}dtdx_iE_i(x)\right)^2} - i\frac{g'}{2}dtdx_iE_i(x) \right] V_i(x).
\end{eqnarray}

For the SU(2) gauge fields:
\begin{eqnarray}
\frac{E^a_{i}(x)-E^a_{i}(x-0)}{dt} &=&
 \sum_j\frac{1}{gdx_id^2x_j}\left(
 {\rm Tr} \left[ i\sigma^a U_i(x) U_j(x+i) U_i^\dagger(x+j) U_j^\dagger(x) \right]
 \right. \nonumber\\
 & &\qquad\left.- {\rm Tr} \left[ i\sigma^a U_j^\dagger(x-j)U_i(x-j)U_j(x+i-j) U_i^\dagger(x)   \right]
 \right) 
  \nonumber\\
 & & - J^{i,a}_{\Phi} - \frac{\partial \Delta {\mathcal L_2}}{\partial W^a_i},
\end{eqnarray}
with
\begin{eqnarray}
U_i(x+0) = \left[\sqrt{1-\sum_a\left(\frac{g}{2}dtdx_iE_i^a(x)\right)^2} + \sum_a i\sigma^a\frac{g}{2}dtdx_iE_i^a(x) \right] U_i(x).
\end{eqnarray}
For hypercharge U(1) and SU(2) gauge fields, Higgs currents are defined respectively, 
\begin{eqnarray}
j^i_{\Phi}  = \left(- \frac{g'}{2dx_i}\right) {\rm Tr} \left[ \Phi^\dagger(x) U_i(x) \Phi(x+i) Z_i(x) i\sigma^3  \right],
\nonumber \\
J^{i,a}_{\Phi}  = \left(- \frac{g}{2dx_i} \right) {\rm Tr} \left[ \Phi^\dagger(x) i\sigma^a U_i(x)\Phi(x+i) Z_i(x) \right].
\end{eqnarray}

In addition to the equations of motion, Gauss' laws can also be derived as derivatives of the lagrangian with respect to temporal components of the gauge fields,
\begin{eqnarray}
\label{equ:gauss}
& &\sum_i\frac{E_i(x)-E_i(x-i)}{dx_i} = \rho^0
,\nonumber \\
& &\sum_i\frac{E^a_i(x)-\bar E^a_{i}(x-i)}{dx_i} = \rho^{a}+ \frac{\partial \Delta {\mathcal L_2}}{\partial W^a_0},
\end{eqnarray}
where,
\begin{eqnarray}
\bar E^a_i(x) = -\frac{1}{2} \sum_b{\rm Tr} \left[ i\sigma^a U^\dagger_i(x) i\sigma^b U_i(x)\right]E^b_i(x)
,\end{eqnarray}
and
\begin{eqnarray}
\rho^0 &=& g' \left[
 \Phi^3 \dot\Phi^0
+\Phi^2 \dot\Phi^1
-\Phi^1 \dot\Phi^2
-\Phi^0 \dot\Phi^3
 \right] 
,\nonumber \\
\rho^{1} &=& g \left[
 \Phi^1 \dot\Phi^0
-\Phi^0 \dot\Phi^1
-\Phi^3 \dot\Phi^2
+\Phi^2 \dot\Phi^3
 \right] 
,\nonumber \\
\rho^{2} &=& g \left[
 \Phi^2 \dot\Phi^0
+\Phi^3 \dot\Phi^1
-\Phi^0 \dot\Phi^2
-\Phi^1 \dot\Phi^3
 \right] 
,\nonumber \\
\rho^{3} &=& g \left[
 \Phi^3 \dot\Phi^0
-\Phi^2 \dot\Phi^1
+\Phi^1 \dot\Phi^2
-\Phi^0 \dot\Phi^3
 \right] .
\end{eqnarray}
The sum of (\ref{equ:gauss}) over the lattice with periodic boundaries further impose constraints on the Higgs charge densities,
\begin{eqnarray}
\label{equ:constraint}
d^3x \sum_x \rho^\beta(x) =0 
,\qquad  {\rm for} \qquad\beta=0,~1,~2,~3.
\end{eqnarray}

The electromagnetic field after symmetry breaking can be calculated by the gauge-invariant combination,
\begin{eqnarray}
A_\mu = \frac{2\sin\theta}{dx_\mu g} \frac{1}{2}{\rm Tr} \left[ i\sigma^3 \frac{\Phi^\dagger(x)}{|\Phi(x)|} U_\mu(x)  \frac{\Phi(x+\mu)}{|\Phi(x+\mu)|} Z_\mu(x)\right]
-\frac{2}{dx_\mu g'\cos\theta} {\rm Im} \left[ V_\mu(x)\right].
\end{eqnarray}

\section{Initial condition}
\label{sec:initial}

\subsection*{Modes and initial charge}

The  Higgs field $\Phi$ and its canonical momenta $\dot \Phi$ are expanded on the lattice as
\begin{eqnarray}
\Phi^\beta(x) = \frac{1}{\sqrt{V}} \sum_p e^{ipx} \frac{1}{\sqrt{2\omega_p}} \eta_p^\beta \sqrt{n_p+\frac{1}{2}}
,\quad
\langle\eta_p^\beta\eta_p^{\beta\dagger}\rangle = 2,
\nonumber \\
\Pi^\beta(x) = \frac{1}{\sqrt{V}} \sum_p e^{ipx} \sqrt{\frac{\omega_p}{2}} \xi_p^\beta \sqrt{n_p+\frac{1}{2}}
,\quad
\langle\xi_p^\beta\xi_p^{\beta\dagger}\rangle = 2,
\end{eqnarray}
where $x_i=s_idx_i$, $p_i=q_i2\pi/(N_idx_i)$ with integer $s_i$ and $q_i$, and $\eta$ and $\xi$ are complex random numbers with Gaussian distributions.
A ``bulk" momentum mode satisfies one of following three criteria:
\begin{eqnarray}
& &(1) \quad 
q_1<(N_1-q_1)\%N_1 
;\nonumber \\
& &(2) \quad 
q_1=(N_1-q_1)\%N_1,~q_2<(N_2-q_2)\%N_2 
;\nonumber \\
& &(3) \quad 
q_1=(N_1-q_1)\%N_1,~q_2=(N_2-q_2)\%N_2,~q_3<(N_3-q_3)\%N_3
,\end{eqnarray}
whereas a ``corner" mode obeys
\begin{eqnarray}
q_1=(N_1-q_1)\%N_1,~q_2=(N_2-q_2)\%N_2,~q_3=(N_3-q_3)\%N_3.
\end{eqnarray}
Therefore, the fields can also be expanded into:
\begin{eqnarray}
& &\Phi^\beta(x) = \frac{1}{\sqrt{V}} \sum_{p \in bulk}  \left(\frac{e^{ipx}}{\sqrt{2\omega_p}} \left[ a_p^\beta+ib_p^\beta\right] \sqrt{n_p+\frac{1}{2}} +h.c \right)
+\frac{1}{\sqrt{V}} \sum_{p \in corner}  \frac{e^{ipx}}{\sqrt{\omega_p}} a_p^\beta  \sqrt{n_p+\frac{1}{2}}
,\nonumber\\
& &\Pi^\beta(x) = \frac{1}{\sqrt{V}} \sum_{p\in bulk} \left(e^{ipx} \sqrt{\frac{\omega_p}{2}} \left[ c_p^\beta+id_p^\beta\right] \sqrt{n_p+\frac{1}{2}} +h.c\right)
+\frac{1}{\sqrt{V}} \sum_{p\in corner}  e^{ipx}\sqrt{\omega_p} c_p^\beta  \sqrt{n_p+\frac{1}{2}}
,\nonumber
\end{eqnarray}
with real numbers $a^\beta_p$, $b^\beta_p$, $c^\beta_p$, $d^\beta_p$ satisfying
\begin{eqnarray}
\langle (a_p^\beta)^2\rangle = \langle (b_p^\beta)^{2}\rangle = \langle (c_p^\beta)^2\rangle = \langle (d_p^\beta)^2\rangle = 1.
\end{eqnarray}
We only initialise unstable modes, whose momenta fulfil  
\begin{eqnarray}
2\sum_i\left(\frac{1-\cos (p_idx_i)}{d^2x_i} \right) < \frac{m_H^2}{2}
,\end{eqnarray}
and these modes will experience a fast growth during Higgs rolling down.
The constraints (\ref{equ:constraint}) are equivalent to $H=0$ with
\begin{eqnarray}
\nonumber
H=
  \left(d^3x\sum_{x} \frac{\rho^0(x)}{g'}\right)^2
+ \left(d^3x\sum_{x} \frac{\rho^1(x)}{g}\right)^2
+ \left(d^3x\sum_{x} \frac{\rho^2(x)}{g}\right)^2
+ \left(d^3x\sum_{x} \frac{\rho^3(x)}{g}\right)^2
.\end{eqnarray}
So before initialising Higgs fields, we implement flow equations along the virtual time $\tau$ to minimise $H$ to zero,  
\begin{eqnarray}
\frac{dx}{d\tau} = -\frac{\partial H}{\partial x}
,\quad
{\rm for ~~ }x= a_p^\beta,~ b_p^\beta,~ c_p^\beta,~ d_p^\beta 
\quad {\rm and} \quad \beta=0,~1,~2,~3
.\end{eqnarray}
We find, by choosing a proper $d\tau$, $H$ can become numerically tiny, if not exactly zero.

\subsection*{Parity}

Under the parity inversion,
\begin{eqnarray}
\Phi(x) \rightarrow \Phi(-x)
,\quad
U_i(x) \rightarrow U^\dagger_i (-x-i).
\end{eqnarray}
We assume the origin point is located at $(\frac{N_1-1}{2},\frac{N_2-1}{2},\frac{N_3-1}{2})$ on the lattice.
So for $x=(s_1,s_2,s_3)$, the opposite site is $-x=(N_1-1-s_1,N_2-1-s_2,N_3-1-s_3)$. The charge conjugation operation is
\begin{eqnarray}
\Phi(x)\rightarrow \Phi^*(x),\qquad U_i(x)\rightarrow U_i^*(x),\qquad V_i(x)\rightarrow V_i^*(x),
\end{eqnarray}
but since the CP-violation arises from a P-breaking term, we will not need to generate an explicitly C-even ensemble of C-conjugate pairs.

\section{CP-violation, $\Delta {\mathcal L_2}$}
\label{sec:l2}

The CP-violating term is
\begin{align}
\Delta {\mathcal L_2}=  -\frac{3\delta_{CP}}{m_W^2} \phi^\dagger\phi \frac{g^2}{64\pi^2}\epsilon^{\mu\nu\rho\sigma}W^a_{\mu\nu}W^a_{\rho\sigma}.
\end{align}
On the lattice, we adopt
\begin{align}
\frac{g^2}{64\pi^2} \epsilon^{\mu\nu\rho\sigma}W^a_{\mu\nu}W^a_{\rho\sigma} \sim \left(-\frac{1}{4\pi^2d^4x}\right)
{\rm Tr} \left[ I_{01}I_{23}+ I_{02}I_{31}+ I_{03}I_{12} \right], 
\end{align}
where
\begin{eqnarray}
I_{\mu\nu} = \frac{1}{8}& & \left[
U_\mu(x) U_\nu(x+\mu) U^\dagger_\mu(x+\nu) U^\dagger_\nu(x)
\right.\nonumber \\
& &+U_\nu(x) U^\dagger_\mu(x-\mu+\nu) U^\dagger_\nu(x-\mu) U_\mu(x-\mu)
\nonumber \\ & &
+U^\dagger_\mu(x-\mu) U^\dagger_\nu(x-\mu-\nu) U_\mu(x-\mu-\nu) U_\nu(x-\nu)
\nonumber \\ & &
+U^\dagger_\nu(x-\nu) U_\mu(x-\nu) U_\nu(x+\mu-\nu) U^\dagger_\mu(x)
\nonumber \\ & &
\left.-h.c
\right].
\end{eqnarray}

Here we list its derivatives with respect to different fields,
\begin{align}
\hspace{-2cm}
\frac{\partial \Delta {\mathcal L_2}}{\partial W_i^a}
=\left(-\frac{3\delta_{CP} g dx_i}{32\pi^2m_W^3 d^4x} \right)\left(K_1[jk] - K_1[kj] - K_2\right),
\end{align}
where
\begin{align}
K_1[jk]=
& ~~~~{\rm Tr}\left[ i\sigma^a U_i(x) \Pi_{0 k}(x+i) U_j(x+i) U^\dagger_i(x+j) U^\dagger_j(x) \right]
\nonumber\\
&+ {\rm Tr}\left[ i\sigma^a U_i(x) U_j(x+i) \Pi_{0 k}(x+i+j) U^\dagger_i(x+j) U^\dagger_j(x) \right]
\nonumber\\
&+ {\rm Tr}\left[ i\sigma^a U_i(x) U_j(x+i) U^\dagger_i(x+j) \Pi_{0 k}(x+j) U^\dagger_j(x) \right]
\nonumber\\
&+ {\rm Tr}\left[ i\sigma^a U_i(x) U_j(x+i) U^\dagger_i(x+j) U^\dagger_j(x) \Pi_{0 k}(x) \right]
\nonumber\\
& - {\rm Tr}\left[ i\sigma^a U_i(x) \Pi_{0 k}(x+i) U^\dagger_j(x+i-j) U^\dagger_i(x-j) U_j(x-j) \right]
\nonumber\\
& - {\rm Tr}\left[ i\sigma^a U_i(x) U^\dagger_j(x+i-j) \Pi_{0 k}(x+i-j) U^\dagger_i(x-j) U_j(x-j) \right]
\nonumber\\
& - {\rm Tr}\left[ i\sigma^a U_i(x) U^\dagger_j(x+i-j) U^\dagger_i(x-j) \Pi_{0 k}(x-j) U_j(x-j) \right]
\nonumber\\
& - {\rm Tr}\left[ i\sigma^a U_i(x) U^\dagger_j(x+i-j) U^\dagger_i(x-j) U_j(x-j) \Pi_{0 k}(x) \right],
\end{align}
and
\begin{align}
K_2=
&~~~~ {\rm Tr}\left[ i\sigma^a U_i(x) \Pi_{j k}(x+i)  U^\dagger_i(x+0)  \right]
\nonumber\\
&+ {\rm Tr}\left[ i\sigma^a U_i(x)  \Pi_{j k}(x+i+0) U^\dagger_i(x+0)  \right]
\nonumber\\
&+ {\rm Tr}\left[ i\sigma^a U_i(x)  U^\dagger_i(x+0) \Pi_{j k}(x+0)  \right]
\nonumber\\
&+ {\rm Tr}\left[ i\sigma^a U_i(x)  U^\dagger_i(x+0)  \Pi_{j k}(x) \right]
\nonumber\\
& - {\rm Tr}\left[ i\sigma^a U_i(x) \Pi_{j k}(x+i)  U^\dagger_i(x-0)  \right]
\nonumber\\
& - {\rm Tr}\left[ i\sigma^a U_i(x)  \Pi_{j k}(x+i-0) U^\dagger_i(x-0)  \right]
\nonumber\\
& - {\rm Tr}\left[ i\sigma^a U_i(x)  U^\dagger_i(x-0) \Pi_{j k}(x-0)  \right]
\nonumber\\
& - {\rm Tr}\left[ i\sigma^a U_i(x)  U^\dagger_i(x-0)  \Pi_{j k}(x) \right],
\end{align}
with
\begin{align}
\Pi_{\mu\nu}(x) = |\phi(x)|^2 I_{\mu\nu}(x).
\end{align}

\begin{align}
\frac{\delta \Delta {\mathcal L_2}}{\delta W_0^a}
=\left(-\frac{3\delta_{CP} g dt}{32\pi^2 m_W^2d^4x}\right) (K_3[ijk]+K_3[jki]+K_3[kij]),
\end{align}
where
\begin{align}
K_3[ijk]=&~~~~ {\rm Tr}\left[ i\sigma^a  \Pi_{j k}(x+0) U_i(x+0)  U^\dagger_i(x) \right]
\nonumber\\
&+ {\rm Tr}\left[ i\sigma^a  U_i(x+0) \Pi_{j k}(x+i+0)  U^\dagger_i(x) \right]
\nonumber\\
&+ {\rm Tr}\left[ i\sigma^a  U_i(x+0)  \Pi_{j k}(x+i) U^\dagger_i(x) \right]
\nonumber\\
&+ {\rm Tr}\left[ i\sigma^a  U_i(x+0)  U^\dagger_i(x) \Pi_{j k}(x) \right]
\nonumber\\
& - {\rm Tr}\left[ i\sigma^a  \Pi_{j k}(x+0) U^\dagger_i(x-i+0)  U_i(x-i) \right]
\nonumber\\
& - {\rm Tr}\left[ i\sigma^a  U^\dagger_i(x-i+0) \Pi_{j k}(x-i+0)  U_i(x-i) \right]
\nonumber\\
& - {\rm Tr}\left[ i\sigma^a  U^\dagger_i(x-i+0)  \Pi_{j k}(x-i) U_i(x-i) \right]
\nonumber\\
& - {\rm Tr}\left[ i\sigma^a  U^\dagger_i(x-i+0)  U_i(x-i) \Pi_{j k}(x)  \right].
\end{align}

\begin{align}
\frac{1}{2}\frac{\partial \Delta {\mathcal L_2}}{\delta\Phi_\beta} k_\beta = \frac{3\delta_{CP}}{4\pi^2m_W^2d^4x}
{\rm Tr} \left[ I_{01}I_{23}+ I_{02}I_{31}+ I_{03}I_{12} \right]  \Phi.
\end{align}


\begin{thebibliography}{*}

\bibitem{CBquench}
  Z.~G.~Mou, P.~M.~Saffin and A.~Tranberg,
  arXiv:1703.01781 [hep-ph].
 
\bibitem{tanmay1}
  T.~Vachaspati,
  Phys.\ Rev.\ Lett.\  {\bf 87} (2001) 251302
  doi:10.1103/PhysRevLett.87.251302
  [astro-ph/0101261].
  
\bibitem{tanmay2}
  C.~J.~Copi, F.~Ferrer, T.~Vachaspati and A.~Achucarro,
  Phys.\ Rev.\ Lett.\  {\bf 101} (2008) 171302
  doi:10.1103/PhysRevLett.101.171302
  [arXiv:0801.3653 [astro-ph]].

\bibitem{Krauss}
  L.~M.~Krauss and M.~Trodden,
  Phys.\ Rev.\ Lett.\  {\bf 83} (1999) 1502
  doi:10.1103/PhysRevLett.83.1502
  [hep-ph/9902420].
  
\bibitem{GB}
  J.~Garcia-Bellido, D.~Y.~Grigoriev, A.~Kusenko and M.~E.~Shaposhnikov,
  Phys.\ Rev.\ D {\bf 60} (1999) 123504
  doi:10.1103/PhysRevD.60.123504
  [hep-ph/9902449].
  
\bibitem{Copeland}
  E.~J.~Copeland, D.~Lyth, A.~Rajantie and M.~Trodden,
  Phys.\ Rev.\ D {\bf 64} (2001) 043506
  doi:10.1103/PhysRevD.64.043506
  [hep-ph/0103231].

\bibitem{Jan1}
  A.~Tranberg and J.~Smit,
  JHEP {\bf 0311} (2003) 016
  doi:10.1088/1126-6708/2003/11/016
  [hep-ph/0310342].

\bibitem{Jan_quench}
  A.~Tranberg, J.~Smit and M.~Hindmarsh,
  JHEP {\bf 0701} (2007) 034
  doi:10.1088/1126-6708/2007/01/034
  [hep-ph/0610096].

\bibitem{wu2}
  A.~Tranberg and B.~Wu,
  JHEP {\bf 1301} (2013) 046
  doi:10.1007/JHEP01(2013)046
  [arXiv:1210.1779 [hep-ph]].

\bibitem{wu1}
  A.~Tranberg and B.~Wu,
  JHEP {\bf 1207} (2012) 087
  doi:10.1007/JHEP07(2012)087
  [arXiv:1203.5012 [hep-ph]].

\bibitem{fermions}
  Z.~G.~Mou, P.~M.~Saffin and A.~Tranberg,
  JHEP {\bf 1506} (2015) 163
  doi:10.1007/JHEP06(2015)163
  [arXiv:1505.02692 [hep-ph]].


\bibitem{juan1}
  A.~Diaz-Gil, J.~Garcia-Bellido, M.~Garcia Perez and A.~Gonzalez-Arroyo,
  Phys.\ Rev.\ Lett.\  {\bf 100} (2008) 241301
  doi:10.1103/PhysRevLett.100.241301
  [arXiv:0712.4263 [hep-ph]].
  
\bibitem{juan2}
  A.~Diaz-Gil, J.~Garcia-Bellido, M.~Garcia Perez and A.~Gonzalez-Arroyo,
  JHEP {\bf 0807} (2008) 043
  doi:10.1088/1126-6708/2008/07/043
  [arXiv:0805.4159 [hep-ph]].


\bibitem{GarciaBellido:2003wd}
  J.~Garcia-Bellido, M.~Garcia-Perez and A.~Gonzalez-Arroyo,
  Phys.\ Rev.\ D {\bf 69} (2004) 023504
  doi:10.1103/PhysRevD.69.023504
  [hep-ph/0304285].


\bibitem{Grasso:2000wj}
  D.~Grasso and H.~R.~Rubinstein,
  Phys.\ Rept.\  {\bf 348} (2001) 163
  doi:10.1016/S0370-1573(00)00110-1
  [astro-ph/0009061].
  
\bibitem{Jan_kappa}
  A.~Tranberg and J.~Smit,
  JHEP {\bf 0608} (2006) 012
  doi:10.1088/1126-6708/2006/08/012
  [hep-ph/0604263].

\bibitem{CPSM}
  T.~Brauner, O.~Taanila, A.~Tranberg and A.~Vuorinen,
  Phys.\ Rev.\ Lett.\  {\bf 108} (2012) 041601
  doi:10.1103/PhysRevLett.108.041601
  [arXiv:1110.6818 [hep-ph]].

\bibitem{tHooft:1974kcl} 
  G.~'t Hooft,
  Nucl.\ Phys.\ B {\bf 79}, 276 (1974).
  doi:10.1016/0550-3213(74)90486-6

\bibitem{sphaleron}
  M.~D'Onofrio, K.~Rummukainen and A.~Tranberg,
  Phys.\ Rev.\ Lett.\  {\bf 113} (2014) no.14,  141602
  doi:10.1103/PhysRevLett.113.141602
  [arXiv:1404.3565 [hep-ph]].
 
  \bibitem{half_Paul}
  A.~Rajantie, P.~M.~Saffin and E.~J.~Copeland,
  Phys.\ Rev.\ D {\bf 63} (2001) 123512
  doi:10.1103/PhysRevD.63.123512
  [hep-ph/0012097].

\bibitem{juan3}
  J.~Garcia-Bellido, M.~Garcia Perez and A.~Gonzalez-Arroyo,
  Phys.\ Rev.\ D {\bf 67} (2003) 103501
  doi:10.1103/PhysRevD.67.103501
  [hep-ph/0208228].

\bibitem{Jan_q}
  J.~Smit and A.~Tranberg,
  JHEP {\bf 0212} (2002) 020
  doi:10.1088/1126-6708/2002/12/020
  [hep-ph/0211243].
  
\bibitem{Skullerud:2003ki} 
  J.~I.~Skullerud, J.~Smit and A.~Tranberg,
  JHEP {\bf 0308}, 045 (2003)
  doi:10.1088/1126-6708/2003/08/045
  [hep-ph/0307094].

\bibitem{mod1}
  K.~Enqvist, P.~Stephens, O.~Taanila and A.~Tranberg,
  JCAP {\bf 1009} (2010) 019
  doi:10.1088/1475-7516/2010/09/019
  [arXiv:1005.0752 [astro-ph.CO]].

\bibitem{mod2}
  B.~J.~W.~van Tent, J.~Smit and A.~Tranberg,
  JCAP {\bf 0407} (2004) 003
  doi:10.1088/1475-7516/2004/07/003
  [hep-ph/0404128].
  
\bibitem{mod3}
  T.~Konstandin and G.~Servant,
  JCAP {\bf 1107} (2011) 024
  doi:10.1088/1475-7516/2011/07/024
  [arXiv:1104.4793 [hep-ph]].
  
\bibitem{inflaton_us}
  Z.~G.~Mou, P.~M.~Saffin and A.~Tranberg,
  In Progress.

\bibitem{Khlebnikov:1988sr} 
  S.~Y.~Khlebnikov and M.~E.~Shaposhnikov,
  Nucl.\ Phys.\ B {\bf 308}, 885 (1988).
  doi:10.1016/0550-3213(88)90133-2

\bibitem{Burnier:2005hp} 
  Y.~Burnier, M.~Laine and M.~Shaposhnikov,
  JCAP {\bf 0602}, 007 (2006)
  doi:10.1088/1475-7516/2006/02/007
  [hep-ph/0511246].

  
  \end{thebibliography}
\end{document}